\documentclass[fleqn,usenatbib,useAMS]{mnras}

\usepackage{graphicx}	
\usepackage{amsmath}	
\usepackage{amssymb}	
\usepackage{multicol}        
\usepackage{bm}		
\usepackage{pdflscape}	

\usepackage[T1]{fontenc}
\usepackage{ae,aecompl}

\usepackage{txfonts}

\title[Faint quenched galaxies in MaNGA]{SDSS-IV~MaNGA: Faint quenched galaxies I- Sample selection and evidence for environmental quenching}

\author[S. J. Penny]{Samantha J. Penny$^{1}$\thanks{Contact e-mail: \href{mailto:samantha.penny@port.ac.uk}{samantha.penny@port.ac.uk}}, Karen L. Masters$^{1}$, Anne-Marie Weijmans$^{2}$, Kyle B. Westfall$^{1}$, \newauthor Matthew A. Bershady$^{3}$,   Kevin Bundy$^{4}$, Niv Drory$^{5}$, Jes\'us Falc\'on-Barroso$^{6,7}$, David Law$^{8}$, \newauthor Robert C. Nichol$^{1}$,  Daniel Thomas$^{1}$, Dmitry Bizyaev$^{9,10}$, Joel R. Brownstein$^{11}$,  \newauthor Gordon Freischlad$^{9}$,  Patrick Gaulme$^{9}$, Katie Grabowski$^{9}$, Karen Kinemuchi$^{9}$,   \newauthor  Elena Malanushenko$^{9}$, Viktor Malanushenko$^{9}$, Daniel Oravetz$^{9}$, \newauthor  Alexandre Roman-Lopes$^{12}$,  Kaike Pan$^{9}$, Audrey Simmons$^{9}$, David A. Wake$^{13,3}$\\
$^{1}$ Institute of Cosmology and Gravitation, University of Portsmouth, Dennis Sciama Building, Burnaby Road, Portsmouth, PO1 3FX, UK\\
$^{2}$ School of Physics and Astronomy, University of St Andrews, North Haugh, St Andrews KY16 9SS, UK\\
$^{3}$ Department of Astronomy, University of Wisconsin-Madison, 475N. Charter St., Madison WI 53703, USA\\
$^{4}$  Kavli Institute for the Physics and Mathematics of the Universe (WPI), The University of Tokyo Institutes for Advanced Study, The University of Tokyo,\\ Kashiwa, Chiba 277-8583, Japan\\
$^{5}$ McDonald Observatory, The University of Texas at Austin, 1 University Station, Austin, TX 78712, USA\\
$^{6}$ Instituto de Astrof\'isica de Canarias, V\'ia L\'actea s/n, E-38205 La Laguna, Tenerife, Spain\\
$^{7}$  Departamento de Astrof\'isica, Universidad de La Laguna, E-38205 La Laguna, Tenerife, Spain\\
$^{8}$ Space Telescope Science Institute, 3700 San Martin Drive, Baltimore, MD 21218, USA\\
$^{9}$ Apache Point Observatory and New Mexico State University, P.O. Box 59, Sunspot, NM, 88349-0059, USA\\
$^{10}$ Sternberg Astronomical Institute, Moscow State University, Moscow\\
$^{11}$ Department of Physics and Astronomy, University of Utah, 115 S. 1400 E., Salt Lake City, UT 84112, USA\\
$^{12}$ Departamento de F\'isica, Facultad de Ciencias, Universidad de La Serena, Cisternas 1200, La Serena, Chile\\
$^{13}$ Department of Physical Sciences, The Open University, Milton Keynes, MK7 6AA, UK\\
}
\date{Last updated 2015 May 22; in original form 2013 September 5}

\pubyear{2015}

\begin{document}
\label{firstpage}
\pagerange{\pageref{firstpage}--\pageref{lastpage}}
\maketitle


\begin{abstract}
Using kinematic maps from the Sloan Digital Sky Survey (SDSS) Mapping Nearby Galaxies at Apache Point Observatory (MaNGA) survey, we reveal that the majority of low-mass quenched galaxies exhibit coherent rotation in their stellar kinematics. Our sample includes all 39 quenched low-mass galaxies observed in the first year of MaNGA. The galaxies are selected with $M_{r} > -19.1$, stellar masses $10^{9}$~M$_{\sun} < M_{\star} < 5\times10^{9}$~M$_{\sun}$, EW$_{H\alpha} <2$~\textrm{\AA}, and all have red colours $(u-r)>1.9$. They lie on the size-magnitude and $\sigma$-luminosity relations for previously studied dwarf galaxies. Just six ($15\pm5.7$~per~cent) are found to have rotation speeds $v_{e,rot} < 15$~km~s$^{-1}$ at $\sim1$~$R_{e}$, and may be dominated by pressure support at all radii. Two galaxies in our sample have kinematically distinct cores in their stellar component, likely the result of accretion. Six contain ionised gas despite not hosting ongoing star formation, and this gas is typically kinematically misaligned from their stellar component. This is the first large-scale Integral Field Unit (IFU) study of low mass galaxies selected without bias against low-density environments. Nevertheless, we find the majority of these galaxies are within $\sim1.5$~Mpc of a bright neighbour ($M_{K} < -23$; or M$_{\star} > 5\times10^{10}$~M$_{\sun}$), supporting the hypothesis that galaxy-galaxy or galaxy-group interactions quench star formation in low-mass galaxies. The local bright galaxy density for our sample is $\rho_{proj} = 8.2\pm2.0$~Mpc$^{-2}$, compared to  $\rho_{proj} = 2.1\pm0.4$~Mpc$^{-2}$ for a star forming comparison sample, confirming that the quenched low mass galaxies are preferentially found in higher density environments.   
\end{abstract}

\begin{keywords}
galaxies: dwarf, galaxies: evolution, galaxies: kinematics and dynamics
\end{keywords}




\section{Introduction}

Identifying the processes which drive the quenching of star formation in the observed galaxy population (and therefore the general decline of star formation in our Universe; \citealt{1996MNRAS.283.1388M}) forms one of the most studied topics in modern extragalactic astrophysics. While it has been recognised for decades that both galaxies which are massive, and those found in higher density environments are more likely to reside on the red sequence and be passive/quenched \citep[][]{1998ApJ...504L..75B,2002MNRAS.334..673L,2003MNRAS.341...54K,2006MNRAS.373..469B,2010ApJ...721..193P,2010MNRAS.404.1775T,2016A&A...586A..23D,2016arXiv160503182D,2016ApJ...818..180C}, disentangling these two correlations has proven difficult. It is now generally recognised that both mass and environment play a role in driving the processes which turn off star formation, such that massive galaxies \citep[in any environment, including voids:][]{2015MNRAS.453.3519P} and galaxies in high density environments (of any mass) are observed to be likely to be quenched  \citep[e.g.][]{2009MNRAS.393.1324B,2010ApJ...721..193P}.

In high density environments (such as large groups or clusters), external processes such as ram-pressure stripping \citep{1972ApJ...176....1G}, tidal 
harassment  \citep{1996Natur.379..613M}  or gas strangulation \citep{1980ApJ...237..692L} are effective in shutting down global star formation. Observations of quenched massive galaxies ($M_{\star} > 3\times10^{10}$~M$_{\odot}$) even in voids demonstrate that internal stellar or halo mass-dependent processes such as supernovae or AGN feedback \citep{2008MNRAS.386.2285C}, and perhaps even processes of secular evolution \citep{2015MNRAS.453.3519P} are able to regulate and shut-off star formation in any environment. However, the degree to which such processes can shape the evolution of low mass galaxies is unknown. AGN feedback might help galaxies maintain quiescence \citep{2006MNRAS.365...11C}, and recently \citet{2016Natur.533..504C} demonstrated that low-powered AGN maintenance-mode feedback (``red geyers'') can heat accreted cold gas, preventing new star formation, even if the AGN feedback is not sufficient to initially quench the galaxy. Simulations have shown that supernova feedback can halt star formation in dwarf galaxies by driving out gas \citep[e.g.][]{1986ApJ...303...39D,2007ApJ...667..170S,2014ApJ...785...58G}, however this ejected gas can cool and re-accrete, resulting in bursty star formation histories in isolated dwarf galaxies \citep[e.g.][]{2007ApJ...667..170S}. Winds driven by star formation can also generate these bursty star formation histories \citep{2015MNRAS.454.2691M,2016ApJ...820..131E}.   

One promising method to isolate the roles in which various processes play in the cessation of star formation, is to identify a sample in which we can be sure that the effect of either environment or galaxy mass on the galaxy's evolution is negligible. In this work we focus our investigation on just the external, or environmental processes by identifying a sample in which internal processes should be unable to completely quench galaxies. Our selection will include low mass galaxies with stellar masses $10^{9} $~M$_{\sun} < M_{\star} < 5\times10^{9}$~M$_{\sun}$. There are few quenched galaxies with low mass found in isolation in the nearby Universe \citep[e.g.][]{2012ApJ...757...85G}, and so it seems clear that they are sufficiently low in mass that internal feedback is inefficient in driving their evolution from star-forming to passive. Galaxies in this mass range are sometimes referred to as dwarf galaxies \citep[e.g.][]{2012MNRAS.419.3167S}, although other studies restrict their dwarf samples to galaxies with $M_{\star} < 10^{9}$~M$_{\odot}$ \citep[e.g.][]{2012ApJ...757...85G}, so in this article we choose to refer to our sample as low mass galaxies rather than dwarf galaxies. 

Despite being the dominant galaxy population by number in groups and clusters, the formation timescale and mechanism of lower mass galaxies, including dwarf spheroidals (dSph, M$_{r} \gtrsim -14$) and dwarf ellipticals (dEs M$_{r}  \gtrsim -19$) are unknown. Like massive galaxies, these lower mass galaxies are observed to follow a clear morphology density relation \citep{1990A&A...228...42B}. This relation is such that dwarf elliptical (dE) and dwarf spheroidal (dSph) galaxies are found primarily in galaxy clusters or at small distances from luminous galaxies in groups, whereas dwarf irregulars (dIrrs) are found in low density regions of the Universe \citep{1990A&A...228...42B}. It therefore seems likely that a large fraction of dE galaxies are late-type galaxies that have been quenched and morphologically transformed by environmental processes \citep{2012ApJS..198....2K,2012ApJ...745L..24J,2013MNRAS.428.2980R}.

The majority of studies examining the origin of lower mass galaxies (excluding the Local Group dwarfs) are restricted to the Virgo Cluster \citep[e.g.][]{2001ApJ...559..791C,2006AJ....132..497L,2011A&A...526A.114T,2014ApJ...783..120T}, with a few studies extending to other clusters, or to the group environment \citep[e.g.][]{2014MNRAS.443.3381P,2015MNRAS.453.3635P}. \citet{2008MNRAS.383..247P}  show that dEs in clusters exhibit a range of star formation histories, with some exhibiting old stellar populations consistent with them being part of the primordial cluster population, whereas other dEs within the same cluster ceased star formation much more recently, likely being a later accreted population. Using IFU spectroscopy of dEs in the Virgo Cluster, \citet{2015MNRAS.452.1888R} show that dEs can have complex star formation histories, with both old and young ($<5$~Gyr) stellar components contributing to their mass. In the Virgo Cluster, these objects show a strong environmental dependence, such that the slow rotators favour the cluster centre, whereas the fast rotators are typically found at large cluster-centric distances. The cluster dwarf galaxy population therefore appears to be a composite of two subpopulations: an old, non-rotating, pressure-supported population found primarily in the cluster core, and a younger, rotationally supported subpopulation, found to larger cluster-centric radii than the pressure-supported systems \citep{2009ApJ...706L.124L,2011A&A...526A.114T}. These fast-rotating systems are likely the low-mass end of the cluster infall population. Many show embedded discy substructure, as revealed by deep imaging and unsharp-masking techniques \citep[][]{2003A&A...400..119D,2003AJ....126.1787G,2006AJ....132..497L}.

Several environmentally-driven mechanisms can quench low-mass star-forming galaxies. In galaxy clusters, ram-pressure stripping by the hot intra-cluster medium can rapidly remove gas from an infalling galaxy in $\sim100$~Myr \citep{1972ApJ...176....1G,1999MNRAS.308..947A,2000Sci...288.1617Q}. However, the effectiveness of this process depends on the gravitational potential of the infalling galaxy, and the size of the galaxy group/cluster, such that low-mass discs falling into massive clusters will be the most strongly stripped. Tidal processes can also drive the evolution of low-mass galaxies. Through frequent high-speed galaxy-galaxy interactions, galaxy harassment transforms the morphology of a galaxy from disc-like to spheroidal, and is again most efficient in the cluster environment \citep{1996Natur.379..613M}. Thus we might expect quenched dEs in groups to exhibit kinematic and morphological differences to those in rich clusters such as Virgo, Coma, and Perseus, making this study of quenched low mass galaxies selected independent of environment or morphology an important addition to the literature. For example, a dwarf galaxy in the core of a dense galaxy cluster will be subject to many more fast galaxy-galaxy interactions than a dE located in e.g. the Local Group. Harassment \citep[e.g.][]{1996Natur.379..613M} is likely more effective at erasing the rotation in the dE's progenitor disc in a rich galaxy cluster vs. a poor galaxy group. 

The kinematics of galaxies provide information on their assembly history and origins, with coherent rotation revealing a more quiescent merger history \citep[e.g.][]{2014MNRAS.444.3357N}, and/or monolithic collapse \citep{1962ApJ...136..748E}, while pressure supported systems can be created via frequent major mergers or collapse via violent relaxation. A rotationally supported system can be transformed into a pressure supported one through galaxy-galaxy interactions via processing including harassment \citep{1996Natur.379..613M}, while infall of cold gas is thought to be able to create new rotationally supported components in galaxies \citep[e.g.][]{2006MNRAS.373..906M}. This rotationally supported component is likely to be kinematically distinct or even counter-rotating from older components. 

It has been known for some time that massive galaxies show a range of rotational properties \citep[e.g.][]{1983ApJ...266...41D,1988A&A...202L...5B}, with S0-Sb galaxies exhibiting more rapid rotation than giant ellipticals \citep[e.g.][]{1982ApJ...256..460K},  which is in strong agreement with more recent work.  For example, nearby massive galaxies have been divided into two classes based on their stellar angular momenta: fast and slow rotators \citep{2011MNRAS.414..888E}. As shown in \citet{1990A&A...239...97B} and \citet{2015ApJ...799..172T} this split extends to the faint galaxy regime ($M_{r} > -19$). However, the kinematics of dwarf galaxies have not been studied in a large or representative sample to date, nor examined with large IFU surveys. In this paper, we aim to construct a sample of low mass galaxies, with no pre-selection on morphology or environment, utilising data from the the Sloan Digital Sky Survey \citep[SDSS,][]{2000AJ....120.1579Y}. Our sample is selected from the Mapping Nearby Galaxies at APO \citep[MaNGA, ][]{2015ApJ...798....7B} multi-object IFU survey, which is in the process of observing a representative sample of $\sim10$k galaxies from the SDSS Main Galaxy Sample \citep{2002AJ....124.1810S}, elected to produce a roughly flat stellar mass distribution with $M_{\star}>10^9$~M$_{\sun}$ (calculated using a \citet{2003PASP..115..763C} Initial Mass Function). We define a low mass galaxy to be any galaxy with a luminosity or stellar mass comparable to, or smaller than, that of the Large Magellanic Cloud ($M_{r} \sim -18.6$, $\sim 3\times 10^{9}$~M$_{\odot}$, \citealt{2002AJ....124.2639V}). To compile our sample, we therefore select all galaxies in the first year of MaNGA observations that are low mass ($<5\times10^{9}$~M$_{\sun}$), low luminosity ($M_{r} > -19$), and have no evidence for ongoing star formation. A second paper (Penny et al., in prep.) will examine the stellar angular momentum of our faint quenched galaxy sample. 

This paper is organised as follows. In Section~\ref{sec:sampsel}, we discuss our sample selection, with a description of the MaNGA survey provided in Section~\ref{sec:manga}, the quenched low mass galaxy selection in Section~\ref{sec:qsamp}, and the local environment of each galaxy in Section~\ref{sec:lenv}. Our results are presented in Section~\ref{sec:results}, including results for two low mass quenched galaxies hosting kinematically distinct cores (Section~\ref{sec:kindes}). We discuss our results in Section~\ref{sec:discuss}, and conclude in Section~\ref{sec:conclude}.  We will apply no environmental or morphological selection to our sample, which is selected from galaxies observed by the Sloan Digital Sky Survey (SDSS) galaxy integral field unit (IFU) survey MaNGA (Mapping Nearby Galaxies at Apache Point Observatory) purely on luminosity (as a proxy for mass) plus indicators of current star formation (a combination of optical colour and emission line strengths). This paper is therefore the first in which the origin of quenched low mass galaxies across a range of environments is studied with a large and representative IFU sample. The only previous IFU studies of low mass galaxies to date have been confined to Virgo \citep{2013MNRAS.428.2980R,2015ApJ...804...70G}, with the exception of \citet{2013MNRAS.428.2980R} who examine IFU spectroscopy for just three dEs selected from the field. Throughout this paper, we assume a cosmology with $\Omega_{\Lambda}=0.7$ and $\Omega_{M}=0.3$, and 
$H_{0} = 100h$~km~s$^{-1}$~Mpc$^{-1}$, unless otherwise noted. 

\section{Sample selection}
\label{sec:sampsel}

\subsection{The MaNGA Survey}
\label{sec:manga}

The IFU data used in this work are taken from the MaNGA survey. MaNGA is a multi-object IFU survey, one of three SDSS-IV projects taking observations using the 2.5~metre Sloan Foundation Telescope \citep{2006AJ....131.2332G} at Apache Point Observatory (APO).  Between 2014 and 2020, MaNGA will obtain IFU spectroscopy for $\sim$10,000 galaxies with $M_{\star} > 1 \times 10^{9}$~M$_{\sun}$, and is constructed to have an approximately flat mass selection (Wake et al., in prep.). There are no cuts in morphology, colour, or environment in the target galaxy selection.

MaNGA utilises the Baryon Oscillation Spectroscopic Survey (BOSS) spectrograph \citep{2013AJ....146...32S}. The BOSS spectrograph provides continuous coverage between 3600~\AA\ to 10300~\AA\, at a spectral resolution $R\sim2000$ ($\sigma_{\rm{instrument}} \sim 77$~km~s$^{-1}$). 17 galaxies are observed at once on a single plate, with IFU fibre bundles ranging in angular diameter between 12~arcsec and 32~arcsec. The fibres have a diameter $\sim2$~arcsec, compared to the typical ground-based seeing of 1.5~arcsec at APO. A 3-position dither pattern with a sub-fibre-diameter offset is used to ensure the gaps between fibres are observed. After dithering, the spatial resolution of the data is $\sim2.2 \mathrm{-} 2.5$~arcsec \citep{2015AJ....150...19L}, though the spaxels of the final datacube are sampled at 0.5~arcsec spacing. Based on a Fourier analysis of the observational PSF in optimal observing conditions, we determined that 0.5~arcsec sampling was required in order to not truncate the k-space modes. 

Complete spectral coverage to 1.5~$R_{e}$ is obtained for the majority of targets, though a subset have coverage to 2.5~$R_{e}$ \citep[see][for more details about the target selection]{2015ApJ...798....7B}. The fibre bundle size is matched to the angular size of the galaxy; the low mass galaxies we are interested in for this study are typically targeted with the 19~fibre bundles.  See \citet{2015ApJ...798....7B} and \citet{2015AJ....149...77D} for more detailed descriptions of the MaNGA survey and instrumentation respectively, and \citet{2015AJ....150...19L} for the MaNGA observing strategy. Analysis of the prototype MaNGA (P-MaNGA) observations are presented in \citet{2015ApJ...804..125L}, \citet{2015MNRAS.449..328W}, and \citet{2015MNRAS.449..867B}.

\subsection{Data reduction and analysis}
\label{sec:dap}

The raw data were processed through the MaNGA data reduction pipeline (DRP), as discussed in detail by Law et al., submitted.  In short, the MaNGA DRP extracts, wavelength calibrates, and flux calibrates \citep[see][]{2016AJ....151....8Y} all fibre spectra obtained in each exposure, combined from both the red and blue arms of the two BOSS spectrographs \citep{2013AJ....146...32S}.  These individual fibre spectra are provided as part of DR13 (Yan et al., in prep) and termed row-stacked spectra (RSS). These individual fibre spectra are then used to construct a datacube with a regular grid of $0\farcs5$ ``spaxels'' and spectral channels that are logarithmically sampled widths of $\Delta\log\lambda = 10^{-4}$.  Our analysis is performed solely on these rectified datacubes.

The analysis of the datacubes is performed using the MaNGA data analysis pipeline (DAP); the development of this pipeline is ongoing and will be described in detail in Westfall et al. (in prep). The primary products of the current DAP are stellar-continuum fits (including stellar kinematic measurements), emission-line properties (such as fluxes, equivalent widths, and kinematics), and spectral-index measurements \citep[e.g., the Lick indices from ][]{1998ApJS..116....1T}. For the purposes of this paper, we are primarily interested in the kinematic data products to establish if the low-mass quenched MaNGA galaxies have a disc origin.

The workhorse module of the MaNGA DAP for the stellar-continuum fitting and stellar kinematics measurements is the Penalised Pixel-Fitting (pPXF) method provided by \citet{2004PASP..116..138C}.  For the results presented here, we have fit all spectra in the datacubes without applying any binning scheme.  We fit the first two moments ($V$ and $\sigma$) of the line-of-sight velocity distribution (LOSVD) only using the empirical MILES spectral library \citep{2006MNRAS.371..703S} as template spectra. The DAP is also run using the solar-metallicity STELIB library \citep{2011MNRAS.418.2785M} and the MUISCAT \citep{2012MNRAS.424..157V} stellar population models, though the MILES library was best able to recover velocity dispersion in testing of the DAP code (Westfall et al. in prep.). After subtracting the best-fitting stellar continuum model from pPXF, the MaNGA DAP fits the strong \ion{O}{iii}, H$\alpha$, \ion{N}{ii}, and \ion{S}{ii} nebular emission lines.  Due to our selection criteria, however, most of the galaxies in our sample are quenched such that they have little to no detectable nebular emission.

A critical component of measuring velocity dispersions (for both the gas and the stars) is the detailed measurement of the spectral resolution of the datacubes.  As discussed by Law et al. submitted, the resolution measurements provided as part of DR13, largely the same as the data products we analyse here, overestimate the spectral resolution by approximately 10~per~cent (which is equivalently an underestimate of the instrumental dispersion).  This has lead to a systematic overestimate of, e.g., the stellar velocity dispersion for our galaxies.  We therefore apply a correction to the velocity dispersion measurements based on this known issue.  This correction is determined by calculating the mean spectral resolution over the wavelength range fitted by pPXF, and subsequently calculating the equivalent quadrature effect of the 10 per cent resolution systematic on the stellar velocity dispersion measurements.  For example, for a spectral resolution of 70~km~s$^{-1}$, the correction for the systematic error in the resolution measurements is $(77^2 - 70^2)^{1/2} = 32.1$ km~s$^{-1}$, which is then applied in quadrature to the MaNGA DAP measurements of the velocity dispersion.

\subsection{Minimum velocity dispersion measurement}
\label{sec:vdispmin}

Dwarf galaxies are low mass galaxies, with stellar velocity dispersions $<100$~km~s$^{-1}$, and galaxies in the luminosity range of our sample ($-19.1 < M_{r} < -17.6$) can have values of $\sigma_{\star} < 30$~km~s$^{-1}$ \citep[e.g.][]{2015MNRAS.453.3635P}. This is well below the MaNGA instrumental resolution of $\sim76$~km~s$^{-1}$ in the region of the spectrum from which we extract galaxy kinematics. We therefore determine the minimum stellar velocity dispersion $\sigma_{\star}$ that we can reliably measure using MaNGA data, and how this varies depending on the S/N of the spectrum by carrying out an idealised Monte Carlo simulation of the ability of pPXF to recover the true velocity dispersion of a test spectrum. A template giant star from the MILES-thin library is selected for this test with a temperature $T_{eff} = 4275$~K, $[Fe/H] = -0.84$, and $\log_{10}g = 1.59$. As cluster dEs are frequently dominated by old, metal poor stellar populations \citep[e.g.][]{2010MNRAS.405..800P}, we selected a K-giant star to match this. The template has a wavelength range 3546~\AA\ to 7405~\textrm{\AA}, identical to the wavelength range fit when measuring the stellar velocity dispersions presented in this paper. This template was convolved with the MaNGA instrumental resolution, then oversampled by a factor of 10. The oversampled spectrum is then convolved with 1000 values of $\sigma_{\star,in}$ between 20~km~s$^{-1}$ and 150~km~s$^{-1}$, and these inputs are used as input spectra for the pPXF. An arbitrary velocity offset is also applied to the oversampled spectrum to better match a true observation of a galaxy redshifted from $z=0$. 

\begin{figure}
 \includegraphics[width= \columnwidth]{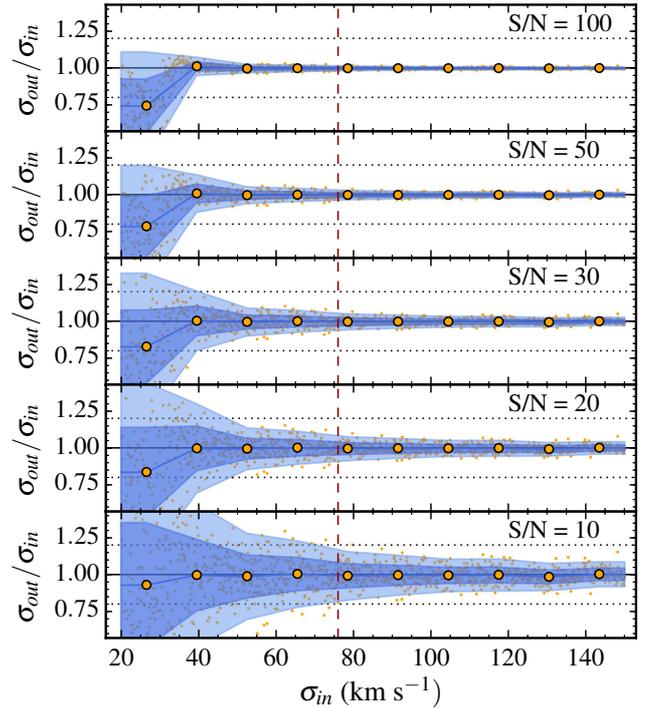}
 \caption{Ratio the velocity dispersion $\sigma_{out}$ recovered by pPXF to $\sigma_{in}$  plotted against $\sigma_{in}$ for five values of S/N (top to bottom: 100, 50, 30, 20 and 10). The small dots are the velocity dispersions calculated as described in Sec.~\ref{sec:vdispmin}, with the mean ratio in bins of width 13~km~s$^{-1}$ shown as large circles. Intervals containing 68 and 95~per~cent (1$\sigma$ and 2$\sigma$) of the values of $\sigma_{out}$ measured by pPXF are shaded dark and light blue respectively. The vertical dashed line in each panel is the MaNGA velocity resolution 76~km~s$^{-1}$ adopted for the simulation, and horizontal lines at $\pm20$~per~cent of $\sigma_{in}$ are provided. \label{fig:sigsim}}
 \end{figure}

The MILES spectrum convolved with the instrumental resolution is then used as a reference template by pPXF, and pPXF run to recover the velocity dispersion of the over-sampled $\sigma$-convolved spectra. This pPXF simulation is set to only recover the first two moments of the kinematics e.g. $V_{\star}$ and $\sigma_{\star}$. This idealised simulation is repeated for signal-to-noise values of 100, 50, 30, 20, and 10, with noise randomly added to the test spectrum at each iteration.  The difference between the true $\sigma_{\star_{in}}$ and the recovered $\sigma_{\star_{out}}$ is then calculated as $\Delta\sigma = \sigma_{out}/\sigma_{in}$ , and this result is plotted in Fig.~\ref{fig:sigsim}.
 
 As can be seen in Fig.~\ref{fig:sigsim}, at high S/N $> 50$, the true velocity dispersion can be recovered to $40$~km~s$^{-1}$, $\sim60$~per~cent of the instrumental resolution. 95~per~cent (2$\sigma$) of the values of $\sigma_{out}$ recovered by pPXF are within 10~per~cent of the original input velocity dispersion $\sigma_{in}$ in these high S/N cases. When the S/N of the input spectrum is decreased, the scatter in measured values of $\sigma_{out}$ increases below the instrumental resolution, such that for a spectrum with S/N 10, the original velocity dispersion can be in error by $\approx20$~per~cent at $\sigma_{in} = 70$~km~s$^{-1}$. Therefore at S/N $\leq10$, only values of $\sigma_{\star}$ similar to, or higher than, the instrumental resolution of $\sim76$~km~s$^{-1}$ can be reliably measured by pPXF for MaNGA data. At high S/N $\gtrsim30$, velocity dispersions to $\sigma_{\star} = 40$~km~s$^{-1}$ can be reliably measured to within $\pm20$~per~cent of their true value, and we therefore adopt this as the minimum value of $\sigma_{star}$ presented in this work. All objects except for one galaxy in our sample have $S/N > 30$ for a binned spectrum containing all spaxels inside $1$~$R_{e}$.  A follow-up paper investigating the stellar angular momentum of the low-mass quenched MaNGA galaxies will examine how the S/N affects the reliability of the measured velocity dispersion in greater detail (Penny et al., in prep.). 

\subsection{Photometry and size determination}

The absolute magnitudes of all the MaNGA galaxies presented in this work are taken from the NASA Sloan Atlas\footnote{\url{http://www.nsatlas.org}}, and were calculated using the sky subtraction technique of \citet{2011AJ....142...31B}. Data in the \textit{u,g,r,i,z} bands are provided by the Sloan Digital Sky Survey \citep[SDSS, ][]{2000AJ....120.1579Y}, while the near-UV magnitudes are provided by \textit{Galaxy Evolution Explorer (GALEX)} telescope \citep{2005ApJ...619L...1M}. 

The colours of the objects are also determined using the values presented in the NASA Sloan Atlas. The magnitudes in all bands are measured using the $r$-band Petrosian aperture, ensuring the same spatial coverage in each photometric band, and thus accurate colour determination. All photometry is corrected for foreground Galactic extinction using the corrections of \citet{1998ApJ...500..525S}, and $k$-corrected to $z=0$ using the \textsc{k-correct} package \citep{2007AJ....133..734B}. The absolute $r$-band magnitudes and colours for the final sample of low mass quenched galaxies are presented in Table~\ref{tab:dprops}.

To provide sizes for our low-mass quenched galaxy sample, the half-light sizes $R_{e}$ (effective radii) of the galaxies are taken from the NASA Sloan Atlas. The NASA Sloan Atlas provides two measurements of $R_{e}$, one taken from a S\'ersic fit to the galaxy's light profile, the other the half-light radius from a non-parametric Petrosian radius. Both sizes are given in Table~\ref{tab:dprops}. In this paper, we choose to use the non-parametric values of $R_{e,Petrosian}$ for the quenched dwarfs when plotting their sizes.  The light profiles of quenched dwarfs are typically found to be fit by multiple-component S\'ersic fits \citep[e.g.][]{2012ApJ...745L..24J}, and thus sizes of the dwarfs may not be determined accurately from a single profile. Thus the Petrosian radius provides a more reliable size measurement for our sample.

\subsection{Quenched low mass galaxy selection}
\label{sec:qsamp}
\begin{table*}
\caption{Basic properties of the MaNGA faint quenched galaxies presented in this paper. The galaxies are selected to have $M_{r} \gtrsim -19$, and Petrosian half-light radii $R_{e} >500$~pc. The redshifts, absolute magnitudes, colours, and sizes of each object are taken from the NASA Sloan Atlas. }
\begin{tabular}{lccccccccc}
\hline
MaNGA ID & Plate & Fibre Bundle & $\alpha$ & $\delta$& $cz$  & $M_{r}$ &  ($u-r$)&  $R_\textrm{e,Petrosian}$ & $R_\textrm{e,S\'ersic}$ \\
& & &(J2000.0) & (J2000.0)&(km~s$^{-1}$)  & (mag) & & (kpc) &  (kpc) \\
\hline
1-38157   & 8083 & 1901  & 03:19:04.24 & +00:56:15.0 & 11265 & $ -18.73 $ & 2.21 & 1.23 & 1.45\\
1-38189   & 8083 & 3701  & 03:20:47.95 & +00:11:08.2 & 11260 & $ -18.46 $ & 2.12 & 1.56 & 1.55 \\
1-38319   & 8083 & 6104  & 03:26:26.27 & -00:18:42.1 & 11278 & $ -18.42 $ & 2.32 & 1.11 & 1.01  \\
1-43680   & 8140 & 3703  & 07:46:18.28 & +41:42:05.9 & 8438 & $ -18.47 $ & 2.22 & 2.13 & 2.45 \\
1-43679   & 8140 & 1901  & 07:46:55.01 & +41:26:41.6 & 8597 & $ -18.57 $ & 2.34 & 1.70 & 3.08 \\
1-217044  & 8247 & 1902  & 09:06:27.82 & +41:38:09.3 & 8191 & $ -18.25 $ & 2.36 & 1.31 & 1.29 \\
1-567184  & 8252 & 1901  & 09:43:05.75 & +48:25:10.9 & 7608 & $ -18.97 $ & 2.20 & 1.27 & 1.33 \\
1-255220  & 8253 & 1902  & 10:31:23.00 & +42:16:37.8 & 6711 & $ -18.65 $ & 2.39 & 0.96 & 1.11 \\
1-277159  & 8256 & 3702  & 10:57:53.04 & +40:51:43.7 & 7526 & $ -18.62 $ & 2.36 & 1.72 & 2.77 \\
1-277154  & 8256 & 1901  & 10:58:30.53 & +40:52:31.9 & 7384 & $ -17.61 $ & 2.24 & 1.21 & 1.41 \\
1-277462  & 8257 & 1901  & 11:00:19.64 & +45:53:24.1 & 6570 & $ -17.66 $ & 2.41 & 0.92 & 1.00 \\
1-256125  & 8451 & 1901  & 11:00:41.70 & +42:31:18.5 & 11465 & $ -18.63 $ & 2.27 & 1.28 & 0.99  \\
1-256457  & 8451 & 1902  & 11:07:07.52 & +42:49:31.8 & 11056 & $ -18.49 $ & 2.31 & 1.51 & 1.16  \\
1-258746  & 8464 & 1901  & 12:24:59.18 & +45:58:10.2 & 7214 & $ -17.59 $ & 2.39 & 1.43 & 1.01 \\
1-456757  & 8480 & 3701  & 12:56:36.01 & +26:54:17.2 & 7895 & $ -17.92 $ & 2.09 & 1.83 & 2.06 \\
12-110746 & 7495 & 1901  & 13:42:09.59 & +27:34:13.3 & 8634 & $ -18.90 $ & 2.28 & 1.00 & 0.93 \\
1-519705  & 8320 & 1902  & 13:42:52.14 & +22:07:37.1 & 8386 & $ -17.76 $ & 2.20 & 0.73 & 0.27 \\
1-235257  & 8325 & 1902  & 14:10:55.18 & +46:03:42.6 & 11992 & $ -18.80 $ & 2.27 & 1.49 & 1.24  \\
1-629695  & 8329 & 1902  & 14:18:55.06 & +45:21:23.9 & 8296 & $ -19.10 $ & 2.37 & 2.08 & 2.45 \\
1-252147  & 8335 & 1902  & 14:32:40.23 & +40:57:15.2 & 5422 & $ -18.34 $ & 2.24 & 0.64 & 0.80 \\
1-322087  & 8552 & 3703  & 15:15:55.85 & +43:53:22.6 & 11795 & $ -18.77 $ & 2.34 & 1.06 & 0.85 \\
12-49536  & 7443 & 1902  & 15:27:57.86 & +42:58:16.3 & 5673 & $ -18.57 $ & 1.96 & 1.70 & 1.68 \\
1-209113  & 8485 & 1902  & 15:42:19.14 & +48:27:56.6 & 11358 & $ -18.55 $ & 2.87 & 1.53 & 0.34 \\
1-322680  & 8315 & 1901  & 15:43:02.18 & +38:39:37.8 & 11201 & $ -18.95 $ & 2.23 & 1.39 & 1.37 \\
1-209078  & 8486 & 3702  & 15:43:05.98 & +48:22:43.7 & 11423 & $ -18.49 $ & 2.07 & 1.53 & 2.00 \\
1-133945  & 8486 & 3703  & 15:55:31.53 & +47:40:38.3 & 5493 & $ -17.16 $ & 2.24 & 1.57 & 1.82 \\
1-133948  & 8486 & 6103  & 15:55:33.91 & +48:01:20.6 & 5851 & $ -18.40 $ & 2.33 & 2.05 & 2.70 \\
1-92638   & 8548 & 1901  & 16:09:08.14 & +47:38:09.9 & 11417 & $ -18.61 $ & 2.08 & 1.03 & 0.98 \\
1-211019  & 8604 & 1902  & 16:28:46.22 & +39:39:21.2 & 9098 & $ -18.91 $ & 2.46 & 1.02 & 1.12 \\
1-211044  & 8312 & 3703  & 16:29:43.71 & +40:00:59.4 & 8281 & $ -18.66 $ & 2.33 & 1.89 & 2.74 \\
1-211098  & 8601 & 3701  & 16:31:24.94 & +39:41:34.6 & 8505 & $ -18.53 $ & 2.50 & 2.26 & 3.37 \\
1-94958   & 8612 & 3702  & 16:53:52.94 & +39:49:46.7 & 10158 & $ -18.44 $ & 2.14 & 1.62 & 2.28  \\
1-136306  & 8606 & 12702 & 17:03:28.78 & +36:26:12.0 & 6946 & $ -18.24 $ & 2.11 & $\ldots$ & $\ldots$\\
1-634477  & 7957 & 1901  & 17:03:34.87 & +36:24:05.5 & 6957 & $ -18.84 $ & 2.29 & 1.42 & 1.79 \\
1-24124   & 7991 & 3703  & 17:14:07.26 & +57:28:38.8 & 8049 & $ -18.54 $ & 2.08 & 2.01 & 2.66 \\
1-24354   & 7991 & 6102  & 17:24:19.64 & +56:52:34.6 & 8249 & $ -18.55 $ & 2.39 & 2.57 & 3.75 \\
1-113525  & 7815 & 1902  & 21:09:43.22 & +11:21:18.9 & 5080 & $ -17.87 $ & 1.92 & 2.33 & 2.74 \\
1-113520  & 7815 & 1901  & 21:10:00.53 & +11:30:38.3 & 5026 & $ -18.29 $ & 2.54 & 0.94 & 1.14 \\
1-115062  & 7977 & 1901  & 22:03:25.29 & +12:40:33.2 & 7811 & $ -18.81 $ & 2.53 & 0.73 & 0.41 \\
\hline
\label{tab:dprops}
\end{tabular}
\end{table*}

\begin{figure}
\includegraphics[width=0.47\textwidth]{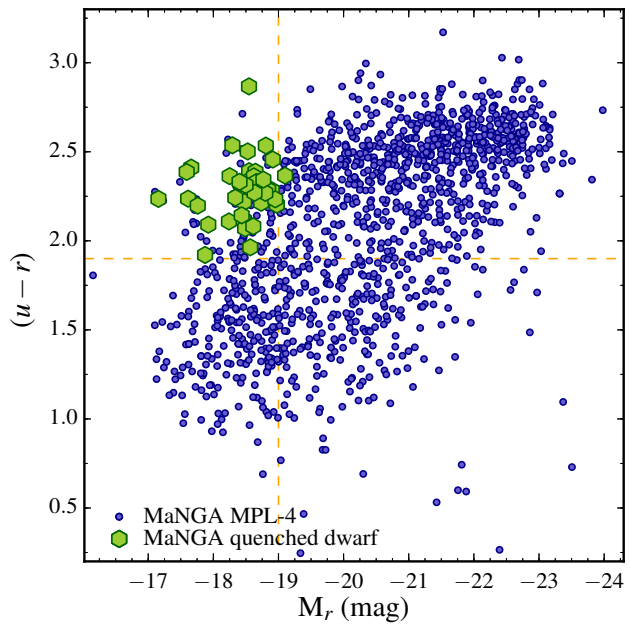}
\caption{$(u- r)$ colour magnitude diagram for galaxies observed during the first six months of the MaNGA survey. The vertical dashed line at $M_{r} = -19$ marks the divide between faint and ``regular'' galaxies, and the horizontal line at $(u-r) = 1.9$ is the lower colour limit for our sample. Faint galaxies with SDSS single fibre spectra consistent with passive stellar populations are plotted as green hexagons. The quenched, faint galaxies have H$\alpha_{EW} < 2$~\AA\ , consistent with little/no ongoing star formation activity. \label{fig:cmd} }
\end{figure}

We draw our sample from the catalogue of galaxies observed during the first year of MaNGA observing (known internally as MPL-4) that have been run through version 1\_5\_1  of the MaNGA DRP,  and also processed by the DAP. These MaNGA datacubes will be released publicly as part of SDSS Data Release 13 (DR13). MPL-4 consists of 1390 galaxy datacubes from the main MaNGA survey observed during the first year of operations, and it is essentially identical to the SDSS-DR13 catalogue. MPL-4 contains 1351 unique targets, with 39 datacubes being repeat observations of the same objects. 

We select galaxies to be quenched using a combination of their luminosity and single-fibre line strengths, with this information taken from the NASA Sloan Atlas. We do not use galaxy colour during the selection process, as the red sequence can contain a mix of genuine non star-forming quenched galaxies, and star-forming galaxies with dust-reddened colours such as edge-on spiral galaxies. Furthermore, at low luminosity, the colours of quenched galaxies  overlap with those of star forming galaxies, such that the red sequence is no longer well defined \citep{2014arXiv1408.5984T}. To identify low-mass objects in the MaNGA survey , we  first select galaxies to have $M_{r} > -19$,  sometimes used as the absolute magnitude dividing dwarf and regular galaxies \citep[e.g.][]{2012ApJ...745L..24J,2014ApJ...783..120T}. A total of 323 galaxies (23~per~cent of MPL-4) are fainter than this magnitude threshold, the majority being low-mass spiral or irregular galaxies.  This absolute magnitude cut restricts the study to galaxies with stellar masses $M_{\star} \lesssim 5 \times10^{9}$~M$_{\sun}$.

Quenched dwarfs are then selected based on the strength of their H$\alpha$ line, with EW$_{H\alpha} <2$~\AA\ as provided in the NASA Sloan Atlas consistent with a galaxy no longer hosting active star formation \citep[e.g.][]{2012ApJ...757...85G}. The galaxies are matched to the NASA Sloan Atlas, which provides these single-fibre measurements. 305 matches are found, though 18 galaxies in our sample are non-matches to the NASA Sloan Atlas. Of the 305 matches,  43 are found to have H$\alpha$ equivalent widths $<2$~\AA\, and do not host ongoing star formation. The majority of the red galaxies with  H$\alpha_{EW} > 2$~\AA\ have a clear edge-on disc morphology, consistent with them being dust-reddened edge-on spirals. 

To check that the remaining 43 galaxies are indeed emission-line free objects, we use a single, all-galaxy spectrum made up of all the stacked spectra in a MaNGA datacube, shifted to the galaxy's line-of-sight velocity, and stacked.  These stacked spectra are checked for emission lines, which removed four star forming disc galaxies. Another object was removed for having an incorrect redshift, being located at $z=0.278$ rather than $z=0.0286$ as listed in the NASA Sloan Atlas, with $M_{r} =  -22.5$. This object (MaNGA 12-180451) furthermore has $\sigma=253\pm18.8$~km~s$^{-1}$, giving a dynamical mass $\gtrsim 10^{11}$~M$_{\sun}$, which places it well outside of the dwarf galaxy regime. 

Binned MaNGA spectra for the 18 non-matches to the NASA Sloan Atlas are then checked to ensure we are not missing any quenched, low mass galaxies in our final sample. An inspection of their single-bin MaNGA spectra (all spaxels binned together) reveals just one of the non-matches to be an emission line free object, and it is therefore kept in our final sample of faint quenched galaxies. The final sample of 39 quenched low mass galaxies have absolute magnitudes $-19.1 < M_{r} < -17.6$, consistent with previous studies of the dynamics of dEs in the Virgo Cluster \citep[e.g.][]{2003AJ....126.1794G,2011MNRAS.413.2665F,2011A&A...526A.114T,2013MNRAS.428.2980R,2015MNRAS.453.3635P}. They furthermore have red colours $(u-r)>1.9$ and lie on the red sequence of the colour magnitude relation (see Fig.~\ref{fig:cmd}), again consistent with them hosting a quiescently evolving stellar population. 

To confirm that these are indeed quenched objects, and not just dusty star forming galaxies, we utilise the strength of the 4000~\AA\ break as an indicator of the very low light-weighted ages of their stellar populations. We follow \citet{2012ApJ...757...85G} who define a quenched galaxy to have EW$_{H\alpha} <2$~\AA\ and a 4000~\AA\ break D$_{n}4000 > 0.6 + 0.1~ \textrm{log$_{10}$}(M_{\star}$). Using this relation, a quenched galaxy in our sample with $M_{\star} \sim 10^{9}$~M$_{\sun}$ would have D$_{n}4000 \gtrsim 1.5$ if it has been quenched for $>500$~Myr.  Most (36/39) objects in this final sample have strong 4000~\AA\ breaks with D$_{n}4000 > 1.5$, indicating they host old stellar populations. One object had D$_{n}4000 = 1.35\pm0.05$, and the remaining two galaxies did not have single-fibre measures of D$_{n}4000$ present in the NASA Sloan Atlas. As none of these three objects exhibit any strong emission, we keep them in our sample of quenched low mass galaxies. This is just a check to confirm these objects are indeed quenched, and a strong value of D$_{n}4000$ is not required for an object to make our final cut. We then confirm that our final galaxies are quenched throughout their structures using MaNGA spectroscopy to provide spatially-resolved emission line strength information. None of the low mass quenched galaxies exhibit emission lines consistent with ongoing star formation, with EW$_{H\alpha} <2$~\AA\  throughout their structures. 

The physical spatial coverage of our sample is variable. 11 objects in our sample are in the primary MaNGA sample, which will target $\sim5000$ galaxies to $\sim1.5$~$R_{e}$. A further 16 objects in our sample are ``colour enhanced'' targets,  selected from an additional $\sim1700$ targets added to the MaNGA primary target list to increase the number of high-mass, blue galaxies, low-mass, red galaxies, and green valley galaxies in the survey. Such objects are relatively rare, but nevertheless trace important stages of galaxy evolution. The galaxies in the colour-enhanced sample are also observed to $1.5$~$R_{e}$. The remaining 11 galaxies are in the secondary MaNGA sample, which contains $\sim3300$ galaxies targeted to $\sim2.5$~$R_{e}$. The targets in the primary sample have $0.017 < z < 0.029$, the targets in the secondary sample have $0.034 < z < 0.040$, and the targets in the colour-enhanced sample have $0.017 < z < 0.030$. See \citep{2015ApJ...798....7B} for a more detailed description of the MaNGA target samples. 
 
Our final sample branches the luminosity gap between IFU surveys of bright early-type galaxies \citep[e.g. SAURON, ATLAS$^{3D}$: ][]{2002MNRAS.329..513D,2011MNRAS.413..813C}, and dE/dSph surveys \citep[e.g][]{2010MNRAS.406.1220W,2011MNRAS.413.2665F,2015MNRAS.453.3635P}. The basic properties of the final sample of low mass quenched MaNGA galaxies are listed in Table \ref{tab:dprops}. They are located at $0.017 < z < 0.040$ ($70~\textrm{Mpc} < D < 160$~Mpc), and have mean velocity dispersions within 1~$R_{e}$ of  $40 \textrm{~km~s$^{-1}$} < \sigma_{e} < 127$~km~s$^{-1}$, spanning those typically found for dEs and low luminosity E/S0 galaxies \citep[see the galaxy sizes presented in][]{2014MNRAS.443.1151N}. Five objects in our sample had values $\sigma_{e}$ that could not be reliably measured due to low S/N $<10$ in the stellar continuum, or having values of $\sigma$ well below the instrumental resolution.

\subsection{Nearest Bright Neighbour}
\label{sec:lenv}

Defining the environment of a galaxy is not simple (see e.g. \citealt{2012MNRAS.419.2670M} for a review on this subject). This situation becomes even more complex when looking at low-mass satellite galaxies including dEs. Quenched dwarf galaxies can have large peculiar velocities relative to the massive galaxies in their host group or cluster \citep[e.g.][]{2001ApJ...559..791C,2001ApJ...548L.139D}. This results in a higher velocity dispersion for the low mass population as a whole than the elliptical galaxies. This increase is thought to be the result of either gravitational interactions with other cluster members (resulting in large peculiar velocities), or due to some low mass galaxies being part of the recent infall population. For example, in the Virgo Cluster, the velocity dispersion of the younger infall dwarf population is higher than that of the original cluster dE population \citep{2009ApJ...706L.124L}. 

The MaNGA survey covers several galaxy clusters, including Virgo, Coma and Leo, and numerous galaxy groups, all of which have very different local densities. Rather than using e.g. 3$^{\textrm{rd}}$ nearest neighbour methods or aperture count methods to describe the local environment of our dE sample, we instead use the distance to their nearest bright/luminous neighbour galaxies. We use this method as quenched dwarf galaxies have been found even in relatively poor galaxy groups  \citep[e.g.][]{2009MNRAS.398..722T}. For each low mass quenched galaxy, we therefore search for the nearest bright galaxy ($M_{K} < -23$) with a measured redshift within 1000~km~s$^{-1}$ of that galaxy. This nearest bright neighbour galaxy is not necessarily the central galaxy of the dwarf's parent halo.

The bright galaxies are selected from the 2MASS redshift survey \citep{2012ApJS..199...26H}, due to its more extended sky coverage than SDSS. We select bright galaxies to have an absolute K-band magnitude M$_{K} > -23$ in a cosmology with $H_{0} = 70$~km~s$^{-1}$, identical to the magnitude cut used by \citet{2012ApJ...757...85G}. This magnitude cut corresponds to a stellar mass $\sim3\times10^{10}$~M$_{\sun}$, the characteristic stellar mass above galaxies are typically dominated by old stellar populations \citep{2003MNRAS.341...54K}. The projected separation and velocity separation of each low-mass galaxy from its nearest bright neighbour is given in Table~\ref{tab:dbright}. 

We use a velocity cut of $\pm1000$~km~s$^{-1}$ to search for each dwarf's nearest bright neighbour galaxy as the velocity dispersion of an individual cluster's dwarf galaxy population can be $>750$~km~s$^{-1}$ \citep[e.g.][]{2001ApJ...559..791C}, due to the large spread in the peculiar velocities of the low mass cluster galaxies. A velocity cut of $<1000$~km~s$^{-1}$ may miss some bright nearest neighbour galaxies due to the broad radial velocity range of dwarfs vs. massive galaxies.  Our velocity cut of $\pm1000$~km~s$^{-1}$ cut ensures we identify the nearest bright neighbour for most dwarfs with large peculiar velocities. We also calculate the local galaxy density of each low-mass galaxy in our sample, as the number of galaxies brighter than $M_{K} = -23$ within a projected distance of 1~Mpc from the dwarf. 

\subsubsection{Star forming comparison sample}

To examine if environment is responsible for quenching low-mass dwarf galaxies, we compare the nearest bright neighbour distances of the quenched faint galaxies to the nearest bright neighbour distance of an approximately magnitude matched sample of dwarf irregular/spiral (dI/dS) galaxies. An identical nearest bright neighbour search is therefore made for a star forming comparison sample.  To compile this comparison sample, a random selection of 39 low luminosity ($M_{r} > -19.1$), blue ($u- r < 1.9$) star forming galaxies is drawn from the same catalogue of observed MaNGA galaxies as the passive sample. The redshift range of the randomly-drawn sample is matched to that found for the low-mass quenched galaxies.  Like the faint quenched galaxies, no environmental constraints are placed on this randomly drawn luminosity-matched comparison sample.  The randomly drawn star-forming sample has $-19.1 < M_{r} < -17.6$, and the nearest bright neighbour separations for the star-forming comparison sample are provided in Fig.~\ref{fig:env}.  
 
\begin{table}
\caption{Separation of each MaNGA faint quenched galaxy from its nearest bright neighbour ($M_{K} < -23$ using $H_{0} = 70$~km~s$^{-1}$ to match \citealt{2012ApJ...757...85G}). The projected separation of the galaxy from its nearest bright neighbour is provided (D$_{proj}$), along with the velocity separation of the two galaxies ($v_{sep}$).}
\begin{center}
\begin{tabular}{lcccc}
\hline
MaNGAID & Nearest Bright Galaxy & $M_{K}$ & $D_{proj}$ & $v_{sep}$ \\
& &(mag)&(kpc)&(km\,s$^{-1}$)\\
\hline
1-38157   &    2MASXJ03191026+0058524 				       & $ -24.25 $ & 98 & 68 \\
1-38189   &    2MASXJ03194561-0004376   			        & $ -24.64 $ & 719 & 186 \\
1-38319   &    2MASXJ03254456-0030307             		        & $ -23.76 $ & 512 & 309 \\
1-43680   &    UGC 04018 					       & $ -25.07 $ & 313 & 135 \\
1-43679   &    UGC 04018  					       & $ -25.11 $ & 139 & 23 \\
1-217044  &    CGCG 209-027  					      & $ -24.46 $ & 341 & 9 \\
1-567184  &    UGC 05145  					      & $ -25.12 $ & 770 & 230 \\
1-255220  &    UGC 05657   					      & $ -24.00 $ & 1077 & 168 \\
1-277159  &    CGCG 213-011  					      & $ -23.81 $ & 51 & 308 \\
1-277154  &    CGCG 213-011					      & $ -23.77 $ & 140 & 166 \\
1-277462  &    UGC 06076 					      & $ -24.38 $ & 45 & 144 \\
1-256125  &    2MASXJ11002009+4227089  			      & $ -23.45 $ & 190 & 173 \\
1-256457  &    2MASXJ11061019+4246476  			      & $ -23.47 $ & 346 & 160 \\
1-258746  &    CGCG 244-011   					            & $ -23.24 $ & 127 & 290 \\
1-456757  &    NGC 4821   					      & $ -24.20 $ & 79 & 850 \\
12-110746 &    IC 4317           				     & $ -23.32 $ & 707 & 6 \\
1-519705  &    CGCG 131-029  					      & $ -24.84 $ & 30 & 125 \\
1-235257  &    2MASXJ14115540+4614170          		      & $ -23.40 $ & 513 & 464 \\
1-629695  &    2MASXJ14174126+4527568          		      & $ -23.23 $ & 347 & 68 \\
1-252147  &    UGC 09386          				      & $ -23.71 $ & 439 & 276 \\
1-322087  &    2MASXJ15220547+4339026          		      & $ -23.75 $ & 2320 & 20 \\
12-49536  &    NGC 5934 					      & $ -24.60 $ & 60 & 110 \\
1-209113  &    2MASXJ15415258+4831191          		      & $ -23.32 $ & 181 & 92 \\
1-322680  &    2MASXJ15432371+3859230           		      & $ -23.24 $ & 652 & 16 \\
1-209078  &    2MASXJ15415258+4831191          		      & $ -23.33 $ & 490 & 157 \\
1-133945  &    UGC 10097 					             & $ -25.11 $ & 184 & 468 \\
1-133948  &    CGCG 250-023 					      & $ -23.84 $ & 119 & 219 \\
1-92638   &    2MASXJ16040373+4745173          		       & $ -23.22 $ & 1702 & 685 \\
1-211019  &    MCG +07-34-074            			      & $ -23.48 $ & 65 & 33 \\
1-211044  &    2MASXJ16293306+3954108       			      & $ -23.21 $ & 170 & 653 \\
1-211098  &    2MASXJ16311927+3944383          		      & $ -23.60 $ & 79 & 766 \\
1-94958   &    2MASXJ16535637+3948459  			       & $ -23.95 $ & 35 & 181 \\
1-136306  &    UGC 10677  					             & $ -24.85 $ & 22 & 156 \\
1-634477  &    UGC 0677  					             & $ -24.86 $ & 26 & 144 \\
1-24124   &    NGC 6338  					       & $ -25.95 $ & 254 & 136 \\
1-24354   &    CGCG 277-042  					       & $ -24.53 $ & 60 & 920 \\
1-113525  &    UGC 11694   					             & $ -24.63 $ & 470 & 15 \\
1-113520  &    CGCG 426-015           				      & $ -23.34 $ & 431 & 14 \\
1-115062  &    NGC 7195                                              & $ -24.26 $ & 34 & 83 \\
\hline
\end{tabular}
\end{center}
\label{tab:dbright}
\end{table}

\section{Results}
\label{sec:results}

\subsection{Local Environment}

We examine the local environment of each dwarf using the distance to their nearest bright neighbour, as described in Section~\ref{sec:lenv}. The majority (37/39) of the quenched low-luminosity MaNGA galaxies have luminous, high-mass neighbours within a projected separation of $1.5$~Mpc and a velocity separation $<1000$~km~s$^{-1}$. Most (36/39) are found within a projected distance $<1$~Mpc from their nearest luminous neighbour, with the smallest projected separation $\sim22$~kpc. The median projected separation for the low-mass quenched galaxies is $184\pm93.1$~kpc, well within the typical virial radius of a galaxy group/cluster ($\approx1.5$~Mpc). For the star forming comparison sample, the median separation is $617\pm157$~kpc. The median velocity separation between a faint quenched galaxy and its nearest bright neighbour is $157 \pm 48$~km~s$^{-1}$, compared to $192\pm57$~km~s$^{-1}$ for the star forming comparison sample.  A difference in local environmental density between the quenched and star-forming samples is also found, with $\rho_{proj} = 8.2 \pm 2.0$~Mpc$^{-2}$ for the quenched galaxies, compared to $\rho_{proj} = 2.1\pm0.4$~Mpc$^{-2}$ for the star forming sample. Thus an environmental difference is seen such that the quenched dwarfs are typically found at higher local projected density than the star-forming comparison sample.

We compare these projected separation results to those for a volume-limited sample of galaxies, which also allows us to check that MaNGA is not biased towards particular environments. We compare this result to the general distribution of low-mass quenched galaxies in the same redshift range as our sample galaxies ($0.017<z<0.040$). To do this, we perform a Monte-Carlo simulation, where a random sample of 39 galaxies with $M_{r} > -19.1$ is randomly drawn from the NASA Sloan Atlas. This random sampling is performed 1000 times. The nearest bright neighbour for each dwarf is found, and the median bright neighbour separation for the random sample is calculated. The general low mass galaxy population has a median projected separation $463\pm24$~kpc, intermediate between that of our low-mass quenched sample, and the randomly-drawn star forming comparison sample. The low-mass quenched galaxies are therefore preferentially found at smaller projected bright-galaxy separations than the general low mass galaxy population, while star-forming low-mass galaxies are typically found at larger separations.

We compare the  projected separations of the quenched and star forming galaxies from their nearest bright neighbours using a two-sided Kolmogorov-Smirnov test. The two-sided K-S test gives $p=0.0042$, and thus we reject the null hypothesis that the low-mass quenched and star forming galaxies are drawn from the same parent population. This test result indicates an environmental dependence, such that the quenched objects are found at smaller separation from their nearest bright neighbour than the star forming comparison sample.  An identical test comparing the velocity separation of the two populations gives $p=0.51$, and the two populations thus have a very similar distribution in velocity separation. The lack of truly isolated dEs in this small sample confirms the importance of local environmental density in shutting off star formation in low mass galaxies. 

\begin{figure}
\begin{centering}
\includegraphics[width=0.44\textwidth]{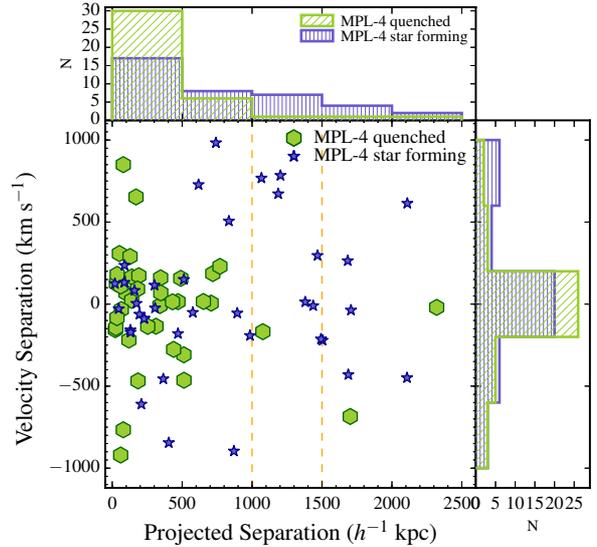}
\caption{Velocity separation of each faint MaNGA quenched galaxy plotted against the separation of each object from its nearest bright neighbour. The low-mass quenched galaxies are shown as green hexagons. The nearest bright neighbour with $M_{K}  < -23$ within $\pm1000$~km~s$^{-1}$ of each galaxy is identified, similar to \citet{2012ApJ...757...85G}, though we use a larger velocity. A comparison sample of randomly drawn, low mass star forming galaxies is included for comparison (blue stars). The vertical lines are located at projected separations of 1~Mpc and 1.5~Mpc. The low-mass quenched galaxies are typically found within 1.5~Mpc of a bright neighbour. }
\label{fig:env}
\end{centering}
\end{figure}

Given the small sample size of 39 dEs presented in this paper, we do not search for environmental trends in the properties of the low mass quenched galaxies, such as size or velocity dispersion differences with respect to local environmental density. By the conclusion of the MaNGA survey in 2020 $\gtrsim200$ early-type faint galaxies will have been targeted for IFU spectroscopy, and clusters including Coma will be targeted as part of the survey.  This will allow for a statistically significant sample of dEs to be examined over a range of environmental density, spanning galaxy groups and clusters.

\subsection{The size-magnitude relation}

\begin{figure}
\begin{center}
\includegraphics[width=0.47\textwidth]{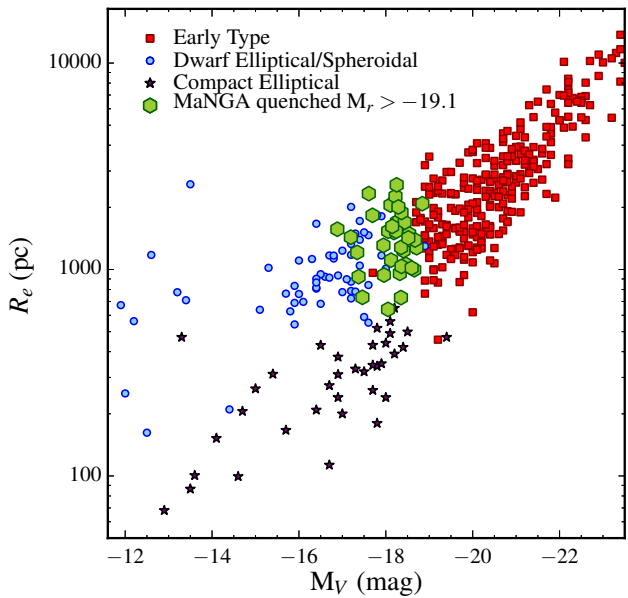}
\caption{The size-magnitude relation for passive stellar systems. The quenched low-mass MaNGA galaxies are shown as green hexagons. The compilation of \citet{2014MNRAS.443.1151N} is shown for comparison, with early-types (primarily taken from the ATLAS$^\textrm{3D}$, \citealt{2011MNRAS.413..813C}) shown as red squares, dwarf ellipticals as blue circles, and compact ellipticals as purple stars.  The MaNGA galaxies follow the same size-magnitude relation as dwarfs galaxies previously examined in the literature, and lie in the region of the diagram where the size-magnitude relation of dEs and cEs diverge.}
\label{sizemag}
\end{center}
\end{figure}

We compare the sizes of the low mass quenched MaNGA galaxies to those of galaxies already studied in the literature. In Fig.~\ref{sizemag}, we plot the size-magnitude relation for early-type stellar systems, as the half-light effective radii $R_{e}$ of the galaxies against their absolute $V$-band magnitude $M_{V}$. The compilation of \citet{2014MNRAS.443.1151N} is plotted for comparison, which is primarily drawn from the ATLAS$^\textrm{3D}$ survey \citep{2011MNRAS.413..813C} for ``normal'' galaxies, with the full range of data sources provided in \citet{2014MNRAS.443.1151N}. All photometry in the \citet{2014MNRAS.443.1151N} catalogue is converted to $V$-band, and thus we perform the same conversion on our SDSS photometry. The $r$-band absolute magnitudes of our dE sample are converted to the $V$-band to match the catalogue of \citet{2014MNRAS.443.1151N}. This is done using the conversion $V = g - 0.55(g-r) - 0.03$ from \citet{2002AJ....123.2121S}. Typical $(g-r)$ colours of the quenched galaxies are $\sim$0.7, so the colour term of this conversion is small.  As the \citet{2014MNRAS.443.1151N} sample is a compilation, with the half-light radii of their compiled galaxy sample measured using multiple techniques and datasets, we do not correct our galaxy sizes for this filter conversion. The majority of the low mass quenched MaNGA galaxies lie on the same region of the size-magnitude diagram as existing dEs, where the relations for dEs and cEs diverge. However, one of the galaxies, MaNGA~1-115062 is unusually small for its luminosity, having a size $R_{e,Petrosian} = 640$~kpc, comparable to that of a compact elliptical.

\subsection{Kinematic maps and velocity profiles}
\label{sec:kinmaps}

\begin{figure*}
\begin{center}
\includegraphics[width=0.88\textwidth]{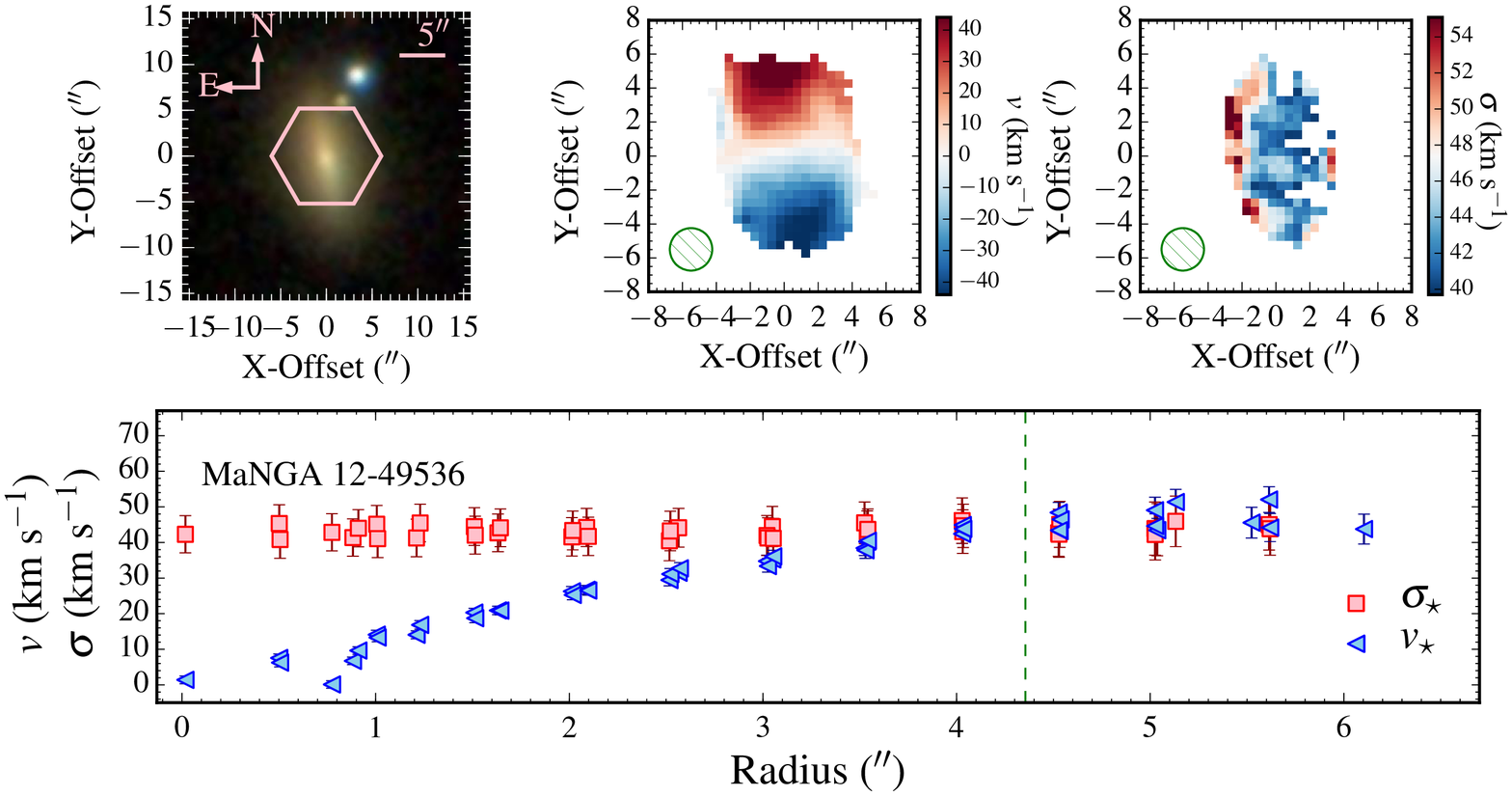}
\includegraphics[width=0.88\textwidth]{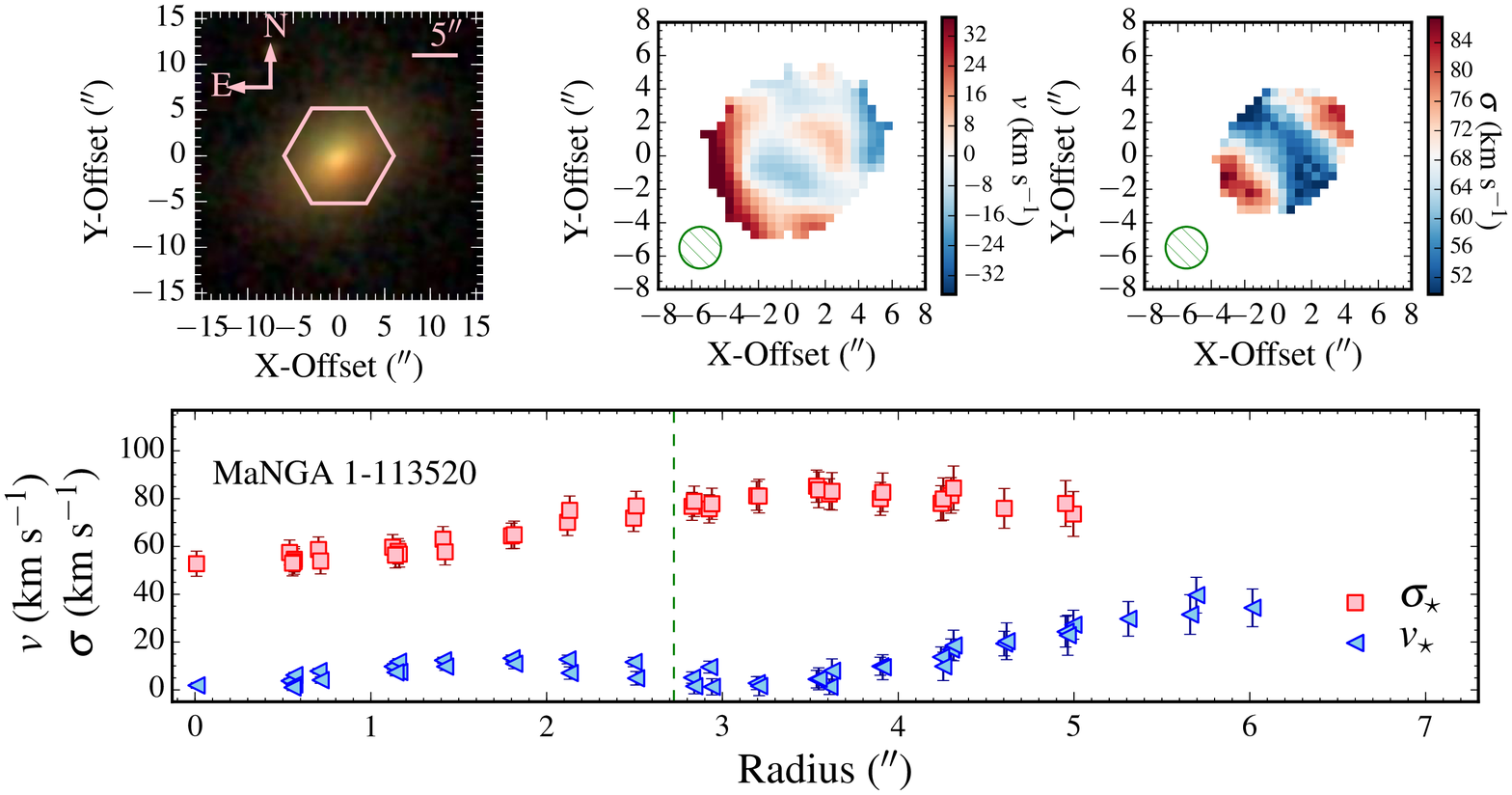}
\caption{The kinematic properties of two example quenched low-mass galaxies in the MaNGA survey. Each plot contains four panels, and only spaxels with values of $\sigma_{\star}$ recoverable at their $S/N$ have been displayed in the maps.  Top left panel: SDSS colour image of each faint quenched galaxy, with the MaNGA IFU field of view overlaid. Top middle panel: observed stellar velocity map to a radius of 5.5~arcsec. Top-right panel: observed stellar velocity dispersion map. Bottom panel: stellar rotation curve extracted along the kinematic major axis. These velocities have not been corrected for beam smearing effects. The typical PSF of a combined MaNGA datacube (2.5~arcsec FWHM) is shown as a green shaded circle. The velocities have not been corrected for the inclination of the galaxy. The majority of objects are rotating, with flat velocity dispersion profiles (e.g. MaNGA~12-49536),  though a few objects exhibit unusual kinematics, including kinematically distinct cores (e.g. MaNGA~1-113520). Kinematic maps for a selection galaxies in our sample are provided in Appendix~\ref{appen1}.}
\label{fig:kinprofs}
\end{center}
\end{figure*}

\begin{table}
\caption{Kinematic information for the quenched MaNGA dwarfs. $b/a$ is the axial ratio of the galaxy taken from the NASA Sloan Atlas, $v_{e,proj}$ is the observed stellar velocity at 1~$R_{e}$ , and $v_{e,rot}$ is the inclination-corrected rotational velocity at the projected radius. The luminosity weighted mean stellar velocity dispersion $\sigma_{e}$ within 1~$R_{e}$ is calculated for each galaxy using a binned datacube containing all spectra within 1~$R_{e}$. The final column provides the S/N of the spectrum from which the kinematics are extracted, with this S/N calculated over the $r$-band region of the spectrum.  \label{tab:kinems}}
\begin{center}
\begin{tabular}{lccccc}
\hline
MaNGA ID & $b/a$ & $v_{e,proj}$ & $v_{e,rot}$& $\sigma_{e}$& $S/N$\\
&& (km~s$^{-1}$) &(km~s$^{-1}$) &(km~s$^{-1}$)&\\
\hline
1-38157   & 0.74 & $ 35 \pm 5.7 $ & $ 52 \pm 7.7 $ & $ 65 \pm 4 $ & 53 \\ 
1-38189   & 0.85 & $ 1 \pm 3.7 $ & $ 2 \pm 3.8 $ & $ <40 $     & 51 \\      
1-38319   & 0.98 & $ 9 \pm 5.0 $ & $ 42 \pm 5.2 $ & $ 103 \pm 4 $ & 49 \\
1-43680   & 0.89 & $ 42 \pm 5.8 $ & $ 94 \pm 6.4 $ & $ 52 \pm 4 $ & 63 \\  
1-43679   & 0.78 & $ 68 \pm 5.6 $ & $ 108 \pm 5.7 $ & $ 78 \pm 4 $& 67 \\  
1-217044  & 0.87 & $ 38 \pm 3.4 $ & $ 77 \pm 3.8 $ & $ 60 \pm 3 $ & 75 \\  
1-567184  & 0.95 & $ 17 \pm 3.6 $ & $ 55 \pm 5.8 $ & $ 74 \pm 3 $ & 111 \\ 
1-255220  & 0.81 & $ 1 \pm 2.3 $ & $ 2 \pm 2.8 $ & $ 83 \pm 3 $ & 129 \\ 
1-277159  & 0.82 & $ 46 \pm 5.5 $ & $ 80 \pm 8.3 $ & $ 70 \pm 3 $  & 88 \\  
1-277154  & 0.80 & $ 29 \pm 4.4 $ & $ 48 \pm 4.7 $ & $ 40 \pm 4 $  & 57 \\  
1-277462  & 0.93 & $  8 \pm 8.3 $ & $ 22 \pm 21.1$ & $ 67 \pm 4 $  & 63 \\  
1-256125  & 0.95 & $ 3 \pm 5.1 $ & $ 11 \pm 8.4 $ & $ 88 \pm 4 $  & 52 \\ 
1-256457  & 0.96 & $ 7 \pm 6.3 $ & $ 27 \pm 7.2 $ & $ 61 \pm 4 $  & 44 \\ 
1-258746  & 0.94 & $ 4 \pm 11.8 $ & $ 11 \pm 13.7 $ & $ 68 \pm 4 $ & 41 \\  
1-456757  & 0.90 & $ 54 \pm 14.9 $ & $ 122 \pm 23.2$ & $ <40 $ & 42 \\       
12-110746 & 0.95 & $ 21 \pm 3.3 $ & $ 70 \pm 5.0 $ & $ 102 \pm 3 $ & 79 \\ 
1-519705  & 0.92 & $ 12 \pm 3.5 $ & $ 30 \pm 6.4 $ & $ 86 \pm 4 $ & 63 \\  
1-235257  & 0.88 & $ 43 \pm 3.5 $ & $ 91 \pm 5.1 $ & $ 57 \pm 4 $ & 57 \\ 
1-629695  & 0.94 & $ 8 \pm 6.6 $ & $ 24 \pm 7.0 $ & $ 73 \pm 3 $ & 89 \\  
1-252147  & 0.76 & $ 15 \pm 1.7 $ & $ 22 \pm 2.0 $ & $ 79 \pm 3 $ & 110 \\ 
1-322087  & 0.92 & $ 21 \pm 2.8 $ & $ 55 \pm 2.8 $ & $ 95 \pm 4 $ & 58 \\ 
12-49536  & 0.63 & $ 45 \pm 3.2 $ & $ 58 \pm 3.4 $ & $ 50 \pm 3 $ & 122 \\ 
1-209113  & 0.97 & $ 22 \pm 4.6 $ & $ 91 \pm 8.9 $ & $ 65 \pm 4 $ & 53 \\ 
1-322680  & 0.95 & $ 22 \pm 3.9 $ & $ 68 \pm 5.0 $ & $ 101 \pm 4 $ & 69 \\
1-209078  & 0.78 & $ 25 \pm 5.2 $ & $ 40 \pm 6.9 $ & $ 51 \pm 4 $ & 46 \\ 
1-133945  & 0.84 & $ 8 \pm 14.2 $ & $ 15 \pm 20.2 $ & $  <40 $ & 37 \\       
1-133948  & 0.76 & $ 8 \pm 6.5 $ & $ 12 \pm 7.6 $ & $ 60 \pm 4$ & 66 \\  
1-92638   & 0.87 & $ 15 \pm 3.8 $ & $ 31 \pm 4.9 $ & $ 70 \pm 4 $ & 41 \\ 
1-211019  & 0.92 & $ 37 \pm 5.5 $ & $ 94 \pm 12.7 $ & $ 115 \pm 3 $ & 88 \\ 
1-211044  & 0.74 & $ 52 \pm 3.9 $ & $ 77 \pm 4.2 $ & $ 72 \pm 4 $ & 65 \\  
1-211098  & 0.87 & $ 44 \pm 7.5 $ & $ 90 \pm 8.9 $ & $ 60 \pm 4 $ & 46 \\  
1-94958   & 0.89 & $ 8 \pm 7.7 $ & $ 18 \pm 8.0 $ & $ 75 \pm 5 $ & 41 \\ 
1-136306  & 0.79 & $\ldots$ & $ \ldots$ & $ <40 $  & 12 \\
1-634477  & 0.85 & $ 44 \pm 5.0 $ & $ 85  \pm 7.3 $ & $ 94 \pm 3 $ & 89 \\  
1-24124   & 0.67 & $ 61 \pm 3.3 $ & $ 83 \pm 3.5 $ & $ 57 \pm 4 $ & 70 \\  
1-24354   & 0.57 & $ 83 \pm 5.6 $ & $ 102 \pm 6.1 $ & $ 72 \pm 4 $ & 66 \\  
1-113525  & 0.78 & $ 13 \pm 6.3 $ & $ 21 \pm 6.4 $  & $ <40 $ & 49 \\       
1-113520  & 0.83 & $ 6 \pm 4.1 $ & $ 11 \pm 5.1 $ & $ 67 \pm 3 $ & 119 \\ 
1-115062  & 0.89 & $ 10 \pm 2.8 $ & $ 22 \pm 4.3 $ & $ 128 \pm 3 $ & 104 \\
\hline
\end{tabular}
\end{center}
\end{table}

We examine the kinematics of the quenched low-mass galaxies using the output from the MaNGA DAP (Section~\ref{sec:dap}). The stellar velocity maps, stellar velocity dispersion maps, radial velocity profiles and $\sigma$ profiles for two faint quenched galaxies are shown in Fig.~\ref{fig:kinprofs} Maps for a selection of low-mass quenched galaxies are provided in Appendix~\ref{appen1}. The radial velocity profiles in Fig.~\ref{fig:kinprofs} have not been corrected for galaxy inclination, nor for the effect of beam smearing. Also included for each object is an SDSS colour image overlaid with the IFU field-of-view. As can be seen in Figure~\ref{fig:kinprofs}, the galaxies exhibit a range of kinematic properties, with some exhibiting clear rotation (e.g. MaNGA~12-49536), while others do not (e.g. MaNGA~1-256125). Within 1~$R_{e}$, the kinematics of most low-mass galaxies are dominated by $\sigma_{\star}$ rather than $v$. 

For each quenched galaxy, the stellar velocity dispersion $\sigma_{e}$ is measured within 1~$R_{e}$ using binned datacubes which two radial bins spanning 0 to 1~$R_{e}$ and 1 to 2~$R_{e}$ respectively. We utilise the measurements from the first bin only, which includes all spectra to 1~$R_{e}$. This binning scheme ensures a high S/N from which each galaxy's stellar velocity dispersion can be robustly measured, with 38/39 dwarfs in our sample having $S/N > 40$ within 1~$R_{e}$. The calculated values of $\sigma_{e}$ for the low-mass quenched galaxies are given in Table~\ref{tab:kinems}. Four galaxies in our sample have values of $\sigma_{e} < 40$~km~s$^{-1}$ despite high S/N spectra,  and their stellar velocity dispersions cannot be accurately measured (see Fig.~\ref{fig:sigsim}). A fifth galaxy, MaNGA  1-136306, has an extremely low $S/N=12$ in its binned spectrum, and its $\sigma_{\star}$ cannot be determined using current MaNGA spectroscopy. The remaining 34 faint quenched galaxies have mean stellar velocity dispersions $40 \textrm{~km~s$^{-1}$} <  \sigma_{e} < 127$~km~s$^{-1}$. 

We also extract the observed stellar velocity $v_{e,proj}$ at the effective radius of each dwarf, along the dwarf's major axis. This  velocity is extracted from their unbinned datacubes, with these values provided in Table~\ref{tab:kinems}. We correct these values of $v_{e,proj}$ for the inclination of each galaxy, and these values of $v_{e,rot}$ also included in Table~\ref{tab:kinems}. The inclination $i$ of each galaxy is calculated as $\cos(i) = b/a $ where $b/a$ is the axial ratio of the galaxy as provided by the NASA Sloan Atlas.  The velocity at one effective radius is then corrected for the inclination of the galaxy as $v_{e,rot} = v_{e,proj}/\sin(i)$.  The low-mass quenched galaxies typically have values of $v_{e,rot} < 100$~km~s$^{-1}$. The majority (23/39) of the galaxies in our sample have $\sigma_{e} > v_{e,rot}$ at 1~$R_{e}$.

\subsection{Gas emission}

While their colours and central H$\alpha$ equivalent widths are consistent with them hosting quenched stellar populations, a fraction of the low-mass 
quenched galaxies exhibit emission lines in their spectra. We inspect their emission line velocity maps, which confirm that the majority of the dEs in our final sample (33/39) are emission-line free objects. The gas velocity maps of these galaxies are consistent with noise maps, and we therefore do not include these maps in this paper. 
 
\begin{figure}
\includegraphics[width=0.22\textwidth]{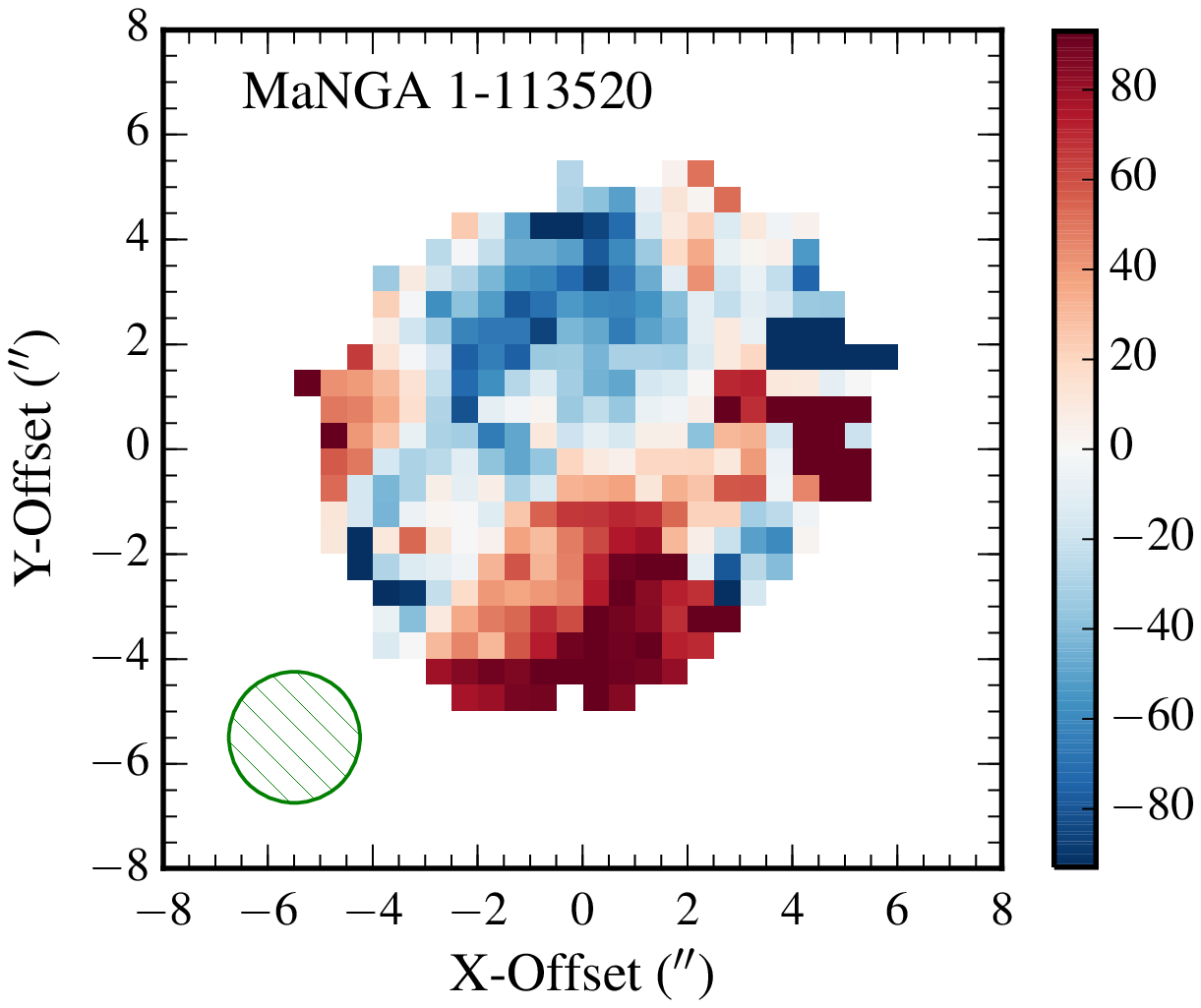}\includegraphics[width=0.22\textwidth]{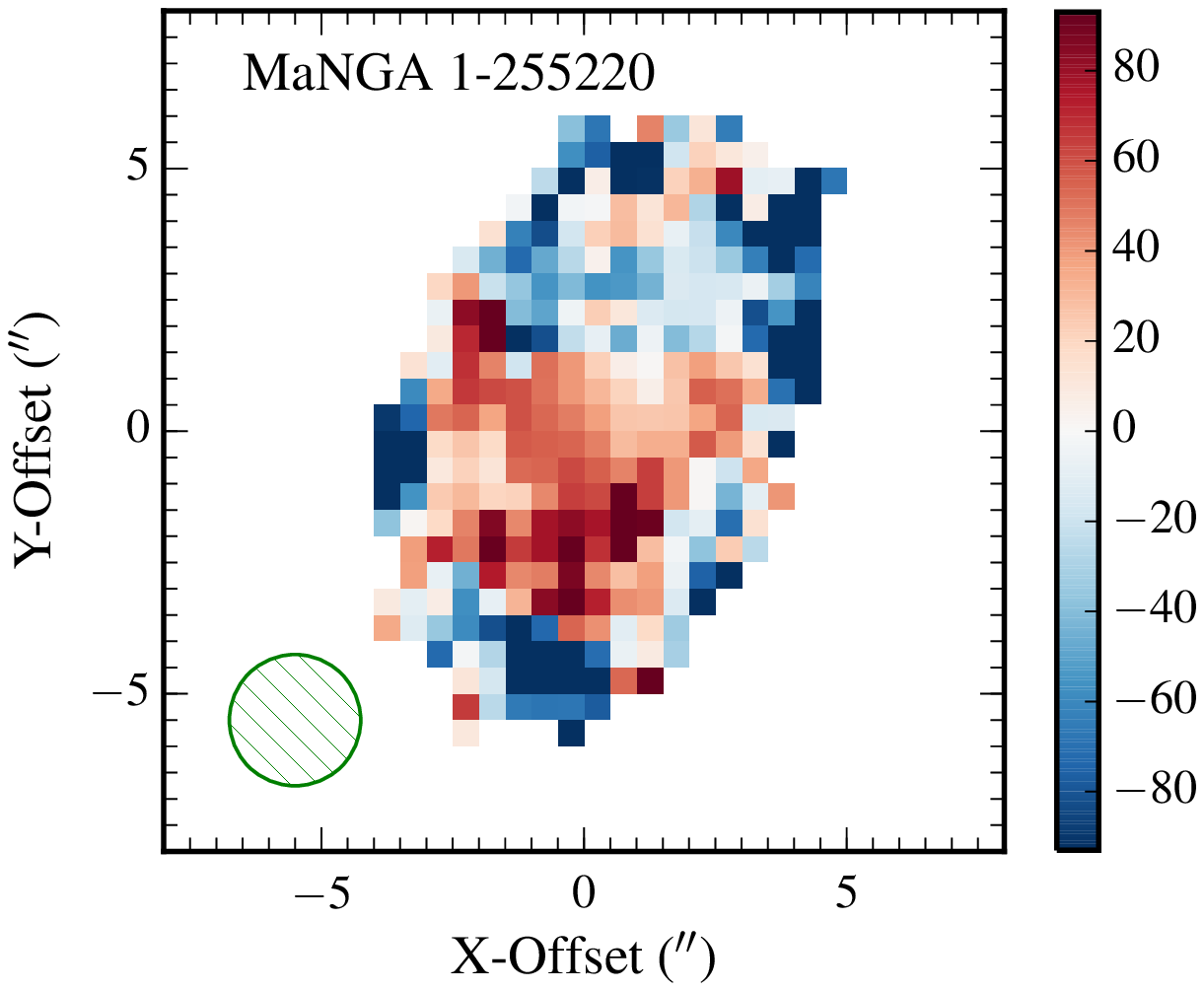}\\
\includegraphics[width=0.22\textwidth]{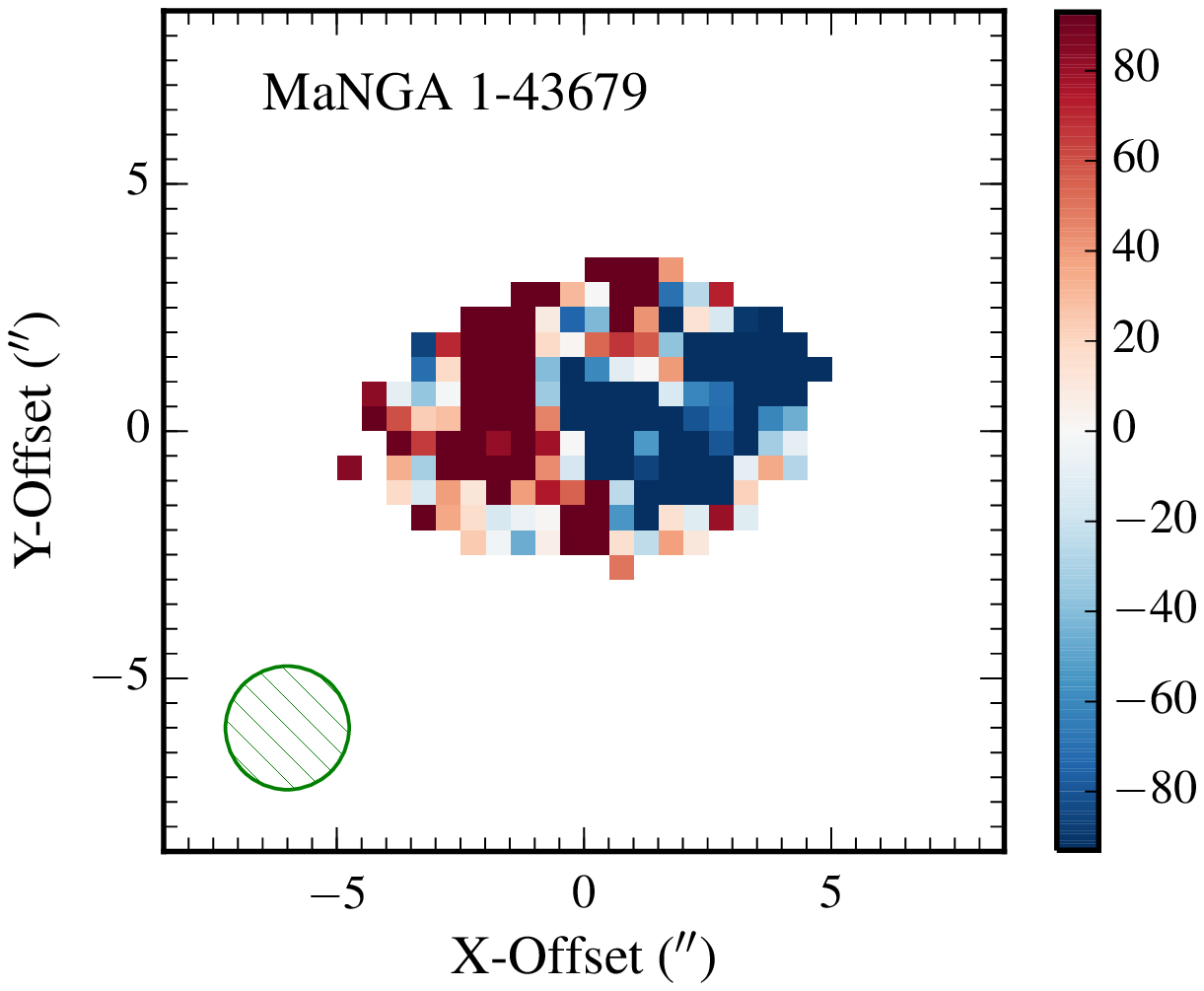}\includegraphics[width=0.22\textwidth]{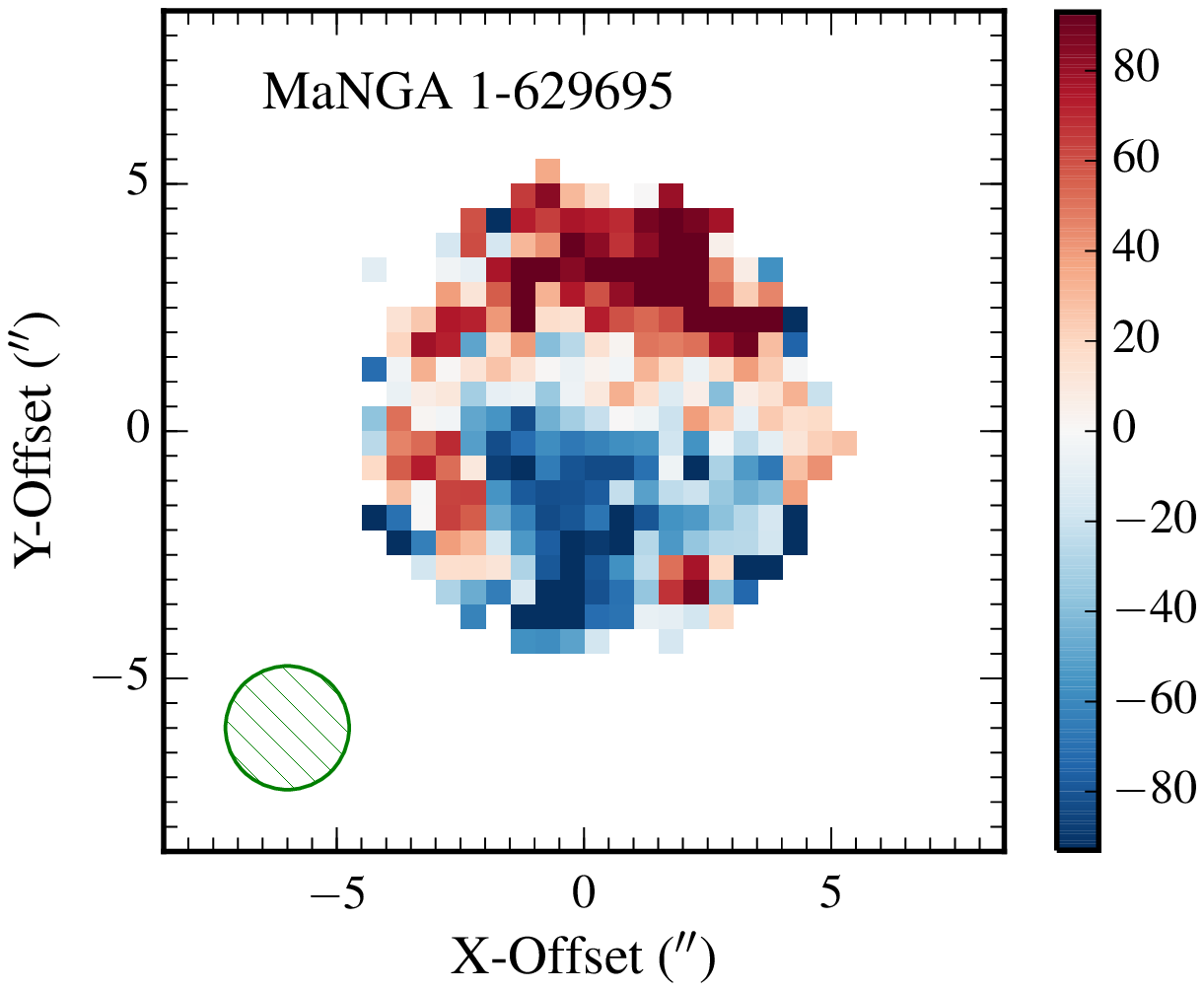}\\
\includegraphics[width=0.22\textwidth]{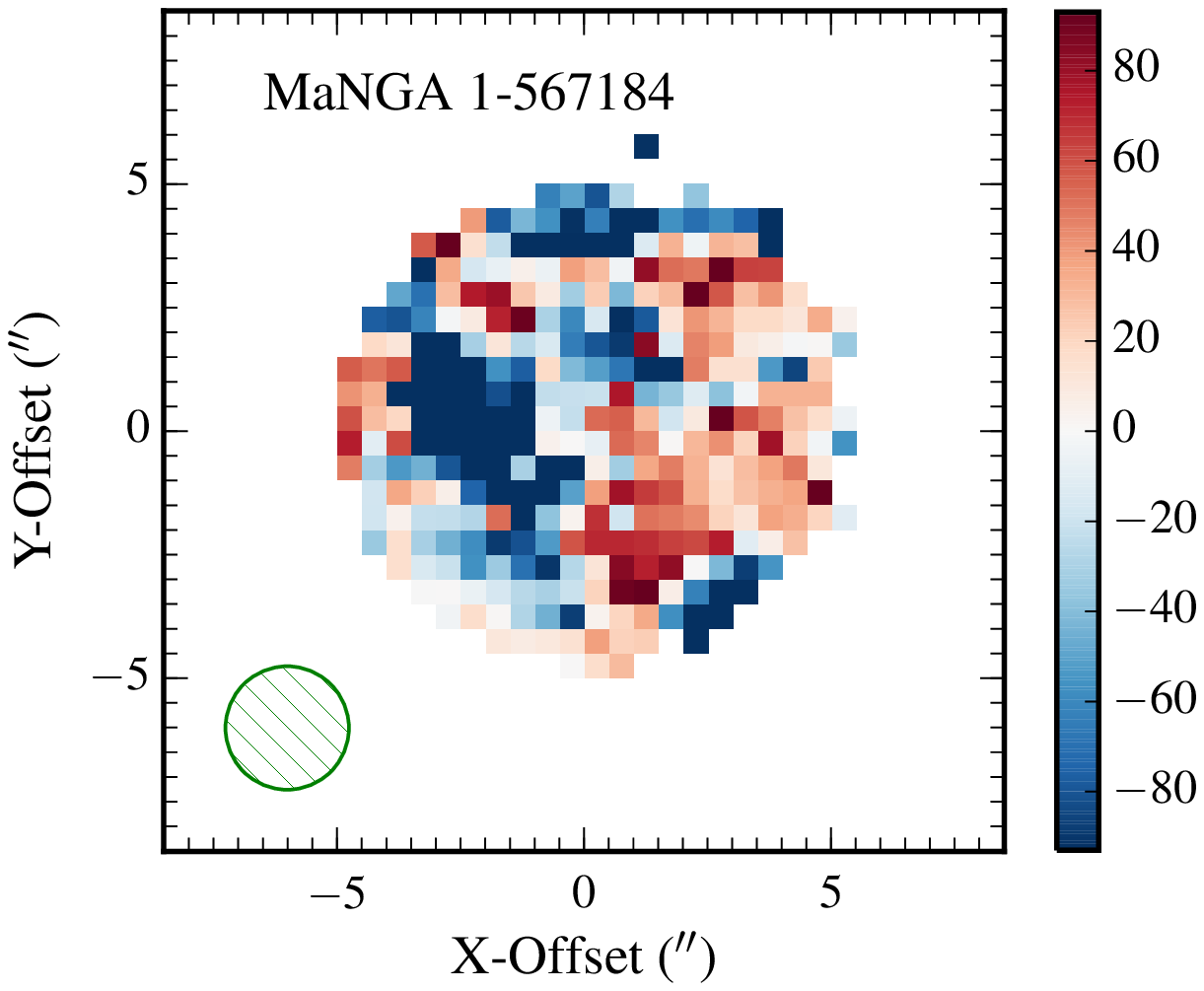}
\includegraphics[width=0.22\textwidth]{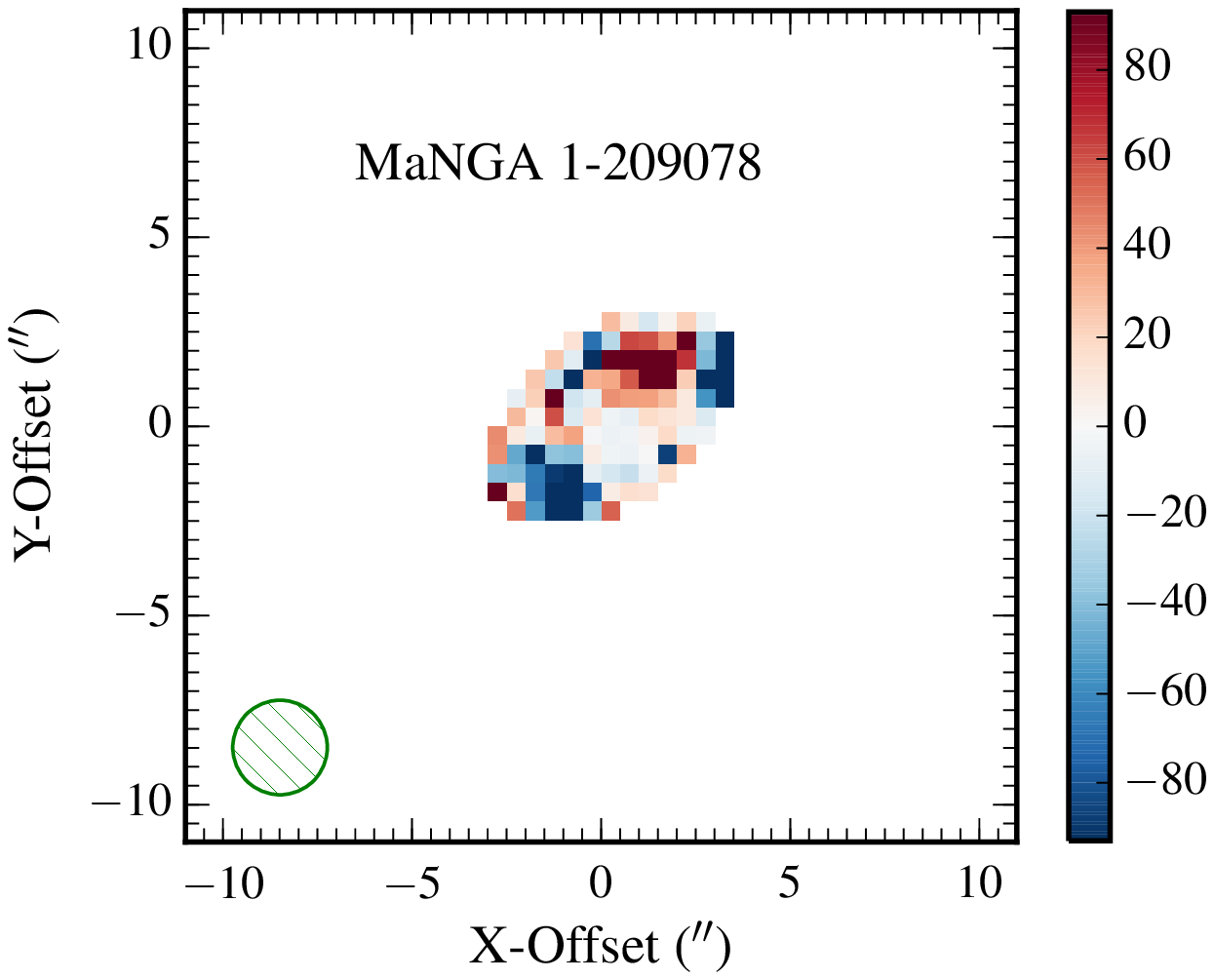}
\caption{Gas-emission line velocity maps for the six faint quenched galaxies in our sample with clear rotation in their ionised gas emission. Their H$\alpha$ emission-line strength is inconsistent with significant ongoing star formation throughout their structures, with EW(H$\alpha$)~$< 2$~\AA.}
\label{fig:gasem}
\end{figure}

However, six galaxies in our sample have noticeable rotation traced by their gas emission, and gas velocity maps for these objects are shown in Fig~
\ref{fig:gasem}. None of these have emission line strengths consistent with any significant star formation throughout their structures. We trace their emission using the average radial velocity of all well-measured emission lines, as provided by the DAP (Section~\ref{sec:dap}). 

For the six objects in which weak gas emission is detected, the rotation of their gas is misaligned with respect to the rotation of their stellar populations. Gas that is kinematically offset from a galaxy's stellar component is a likely indicator of accretion \citep[e.g.][]{2011MNRAS.417..882D}, though in an axisymmetric potential the gaseous component will relax over time, and may be misaligned by either $0^{\circ}$ or $180^{\circ}$. MaNGA~1-113520 (which exhibits a kinematically distinct core), MaNGA~1-209078,  MaNGA~1-629695, and MaNGA~1-567184  have their gas rotation is offset from their stellar rotation by $<180^{\circ}$. MaNGA~1-255220, which exhibits a kinematically distinct core in its stellar component, only exhibits gas emission in its inner region. The rotation of this component is offset by $180^{\circ}$ to the stellar rotation of the galaxy. The remaining galaxy, MaNGA~1-43679  has it ionised gas component rotation offset by  $180^{\circ}$ from its stellar rotation.

\subsection{The $\sigma$-luminosity relation}

\begin{figure}
\begin{center}
\includegraphics[width=0.47\textwidth]{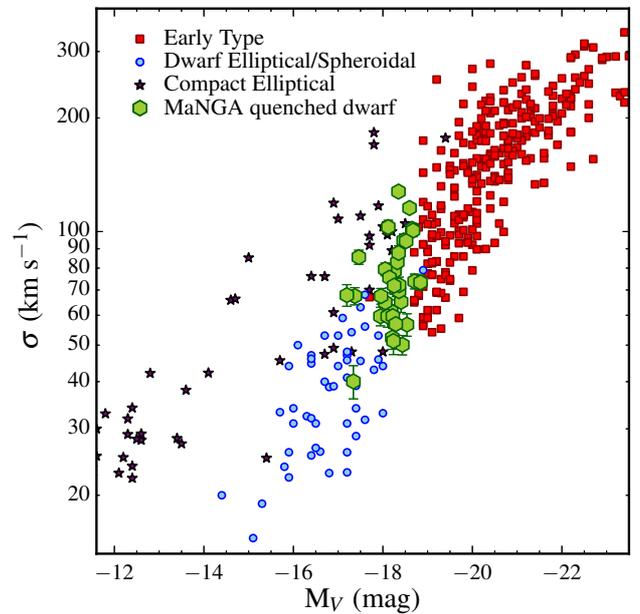}
\caption{The $\sigma$-luminosity relation for passive stellar systems. The quenched MaNGA galaxies are shown as green hexagons. The compilation of \citet{2014MNRAS.443.1151N}  is also plotted for comparison, and the symbols have the same meaning as in Fig.~\ref{sizemag}. The majority of the faint MaNGA galaxies with $(u-r) > 1.9$ have values of $\sigma$ comparable to those of dEs previously examined in the literature. However, some have $\sigma$ values $>100$~km~s$^{-1}$, similar to those typically measured for cEs \citep{2015ApJ...804...70G}.}
\label{siglum}
\end{center}
\end{figure}

The $\sigma$-luminosity relation for quenched stellar systems is show in Fig.~\ref{siglum}, with the MaNGA galaxies plotted as green hexagons. We plot the values of $\sigma_{e}$ as calculated in Section~\ref{sec:kinmaps}. Again, the compilation of \citet{2014MNRAS.443.1151N}  is plotted for comparison, using the same symbols as Fig.~\ref{sizemag}. The majority of the MaNGA galaxies  have velocity dispersions comparable to those already measured in the literature (typically for Virgo dEs), with $\sigma_{e} < 100$~km~s$^{-1}$. However, five have velocity dispersions higher than this, not typical $\sigma$ for dwarf galaxies. The smallest object in our sample (MaNGA~1-115062, $R_{e}  = 676$~pc) lies in the cE region of the diagram. This galaxy exhibits the highest velocity dispersion of any galaxy in our sample, with $\sigma_{e} = 128$~km~s$^{-1}$. This small size and high velocity dispersion, combined with its low luminosity ($M_{r} = -18.8$), makes it a candidate cE galaxy, else an extremely low-luminosity elliptical galaxy.  

\subsection{Kinematically decoupled cores}
\label{sec:kindes}

\begin{figure*}
\begin{center}
\includegraphics[width=0.43\textwidth]{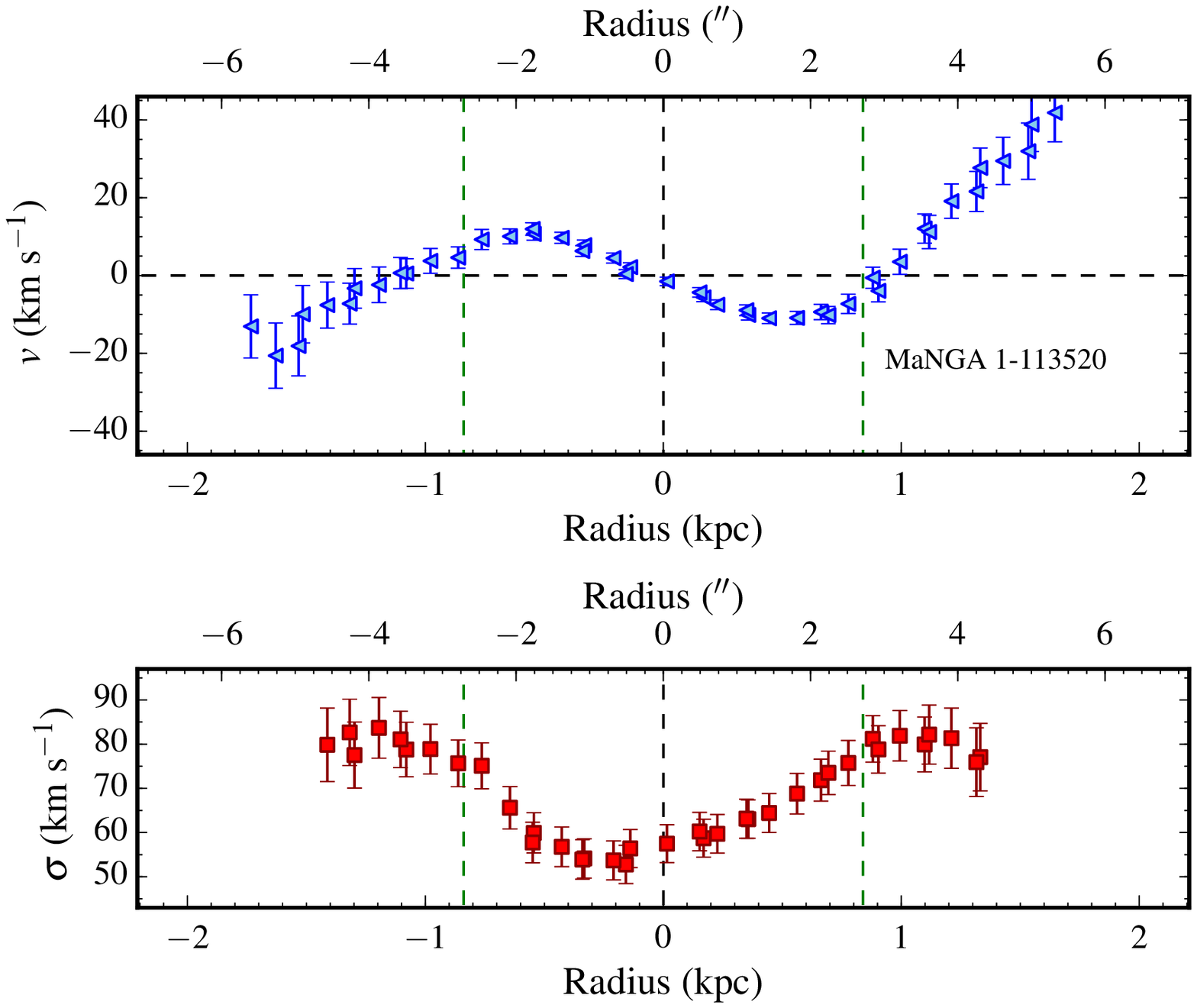}
\includegraphics[width=0.43\textwidth]{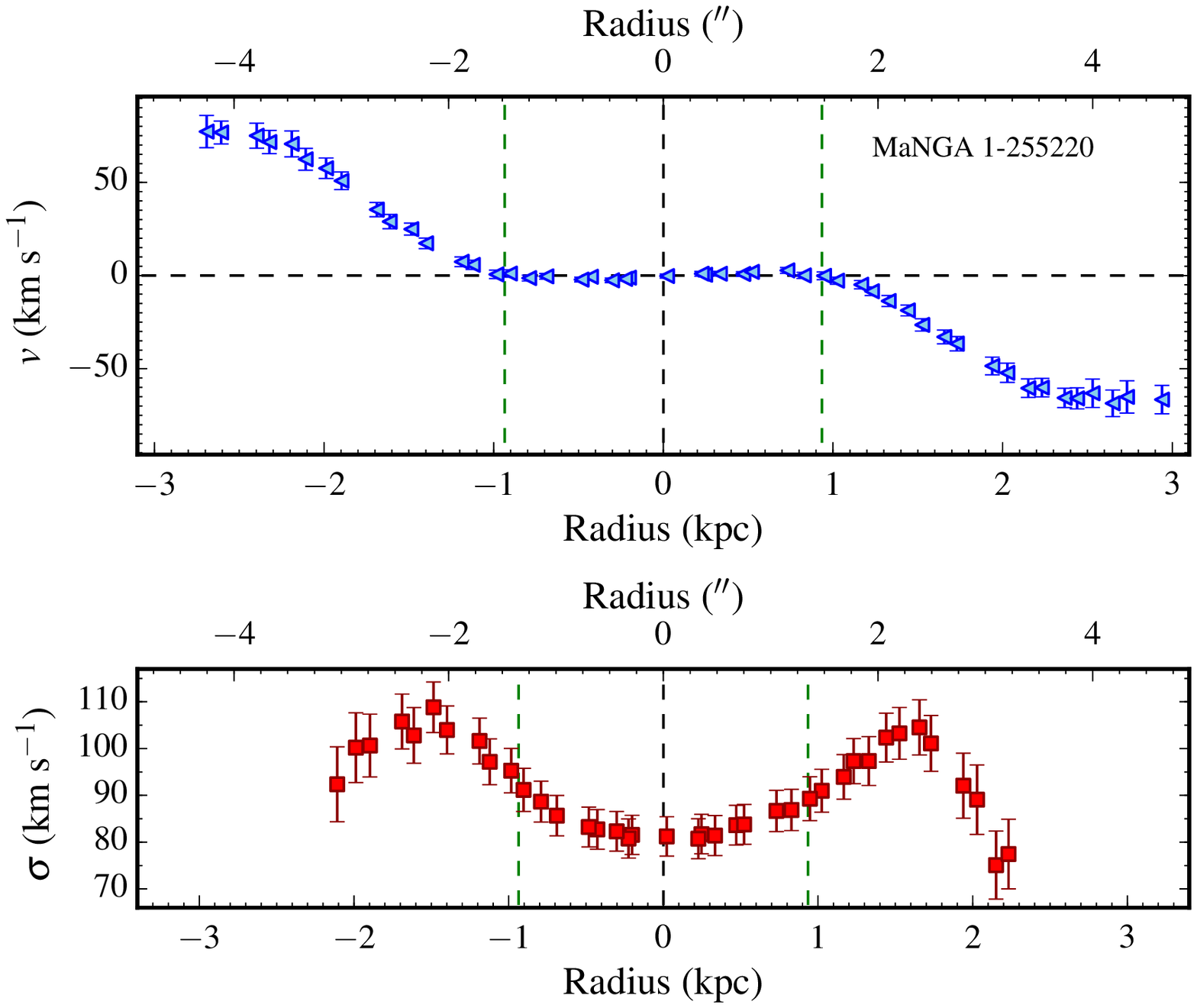}
\caption{Radial velocity (top panel) and stellar velocity dispersion profiles (bottom panel) for two passive faint MaNGA galaxies with kinematically-distinct cores. The vertical dashed lines are located at $R=0$ and $\pm1$~$R_{e,Petrosian}$. The kinematics are taken from each dwarf's datacube, along a 1~arcsec wide aperture to mimic a typical long-slit observation. The aperture is aligned with the dwarf's major axis. The radial velocity profiles for both galaxies invert, and two $\sigma$ 
peaks are clearly seen, indicating the presence of counter-rotating discs in both galaxies.}
\label{fig:kindist}
\end{center}
\end{figure*}

Two of the faint passive galaxies have kinematically distinct cores (KDCs) as revealed by their radial velocity and velocity dispersion maps:  MaNGA 1-113520 and MaNGA 1-255220. They have absolute magnitudes $M_{r} = -18.29$ and $M_{r} = -18.65$ respectively, and have smooth, flattened morphologies with $b/a = 0.75$ and $0.64$ respectively. The line-of-sight stellar velocity curve and velocity dispersion profiles of both galaxies are shown in Figure~\ref{fig:kindist}. The profiles are extracted along a 1~arcsec wide strip centred along each object's major photometric axis, to mimic a typical long-slit observation.  

A  counter-rotating KDC is clearly seen in the stellar rotation curve of MaNGA 1-113520, where a sharp velocity turnover is observed at a radius of 0.5~kpc. MaNGA 1-255220 does not exhibit any significant rotation at radii $< 2.5$~arcsec, $\sim1.2$~kpc, at which the velocity rapidly increases consistent with disc-like rotation. 

Both galaxies also exhibit unusual velocity dispersion profiles, such that the stellar $\sigma$ is lowest at the centre of the galaxy, with the velocity dispersion profiles furthermore flat in the central regions. However,  $\sigma_{\star}$ increases rapidly by $\approx30$~km~s$^{-1}$ with radius at $R > 500$~pc in both galaxies, with $\sigma$ peaks equidistant from each galaxy's centre. These $\sigma$ peaks are offset from the radial velocity turnover by $\sim500$~pc in both galaxies. In addition to their unusual stellar kinematics, both dwarfs exhibit an ionised gas component, which is counter-rotating to their stellar component in both cases. The properties and local environments of the two dwarfs are described in more detail below. 

\subsubsection{MaNGA~1-113520}
\label{sec:kdet}

At $M_{r} = -18.29$, with $R_{e} = 0.94$~kpc and $\sigma_{e} = 67\pm3$~km~s$^{-1}$, MaNGA~1-113520 falls on the regions of the size-magnitude and $\sigma$-luminosity relation typically occupied by dE galaxies. The galaxy exhibits a clear 180$^{\circ}$ twist in its stellar kinematics at $R = 2''$, within 1~$R_{e}$. We therefore assume that the KDC is due to the presence of two counter-rotating discs. The galaxy has a disc-like morphology, and is listed as an S0 galaxy in the NASA Extragalactic Database. It is located at a projected distance of 431~kpc and $v= 14$~km~s$^{-1}$ from the nearest bright galaxy. All galaxies listed in the NASA Extragalactic Database within a projected radius of 1~Mpc from MaNGA~1-113520 are within $v =400$~km~s$^{-1}$, with a velocity dispersion of $\sigma_{group} = 160$~km~s$^{-1}$. The galaxy exhibits weakly-ionised gas emission, and this gas is kinematically misaligned from the stellar rotation. This misaligned ionised gas may be the result of the accretion event responsible for the formation of the kinematically-distinct stellar components. Alternatively, this component could be an ionised wind, similar the ``red geyser'' maintenance-mode feedback identified in \citet{2016Natur.533..504C}.     

\subsubsection{MaNGA~1-255220}

MaNGA~1-255220 has $M_{r} = -18.65$, $R_{e} = 0.96$~kpc and $\sigma_{e} = 83\pm3$~km~s$^{-1}$, and has a inclined edge-on disc morphology. It is remarkably isolated for a quenched low mass galaxy, and its nearest bright neighbour galaxy (UGC 05657, $M_{K} = -24$) is located at a projected separation of 1.53~Mpc. In its central regions, MaNGA~1-255220 exhibits weak H$\alpha$, indicating the presence of ionised gas. This ionised gas is rotating with an offset of 180$^{\circ}$ to the outer stellar component of the galaxy. This suggests the two-$\sigma$ structure seen in the stellar velocity dispersion map of MaNGA~1-255220 is the result of gas accretion, which resulted in the formation of two counter-rotating stellar discs.

\section{Discussion}
\label{sec:discuss}

\subsection{The local environment of low-mass quenched galaxies}

The faint quenched galaxies we identify in this work are typically located within $\sim1.5$~Mpc and $\pm1000$~km~s$^{-1}$ of an $M_{K} = -23$ or brighter galaxy, with a mean separation of  $368\pm74$~kpc, and a median separation $184\pm93$~kpc. The majority (29/39) are within 500~kpc of a galaxy with $M_{K} < -23$ (corresponding to a stellar mass $\sim3\times10^{10}$~M$_{\sun}$). Only 2 low mass galaxies have a separation $>1.5$~Mpc from their nearest bright neighbour and might therefore be considered isolated galaxies. As can be clearly seen in Fig.~\ref{fig:env},  this is in contrast to a randomly drawn mass-matched sample of low-mass star forming galaxies, which have a mean separation of $857\pm125$~kpc, and a median separation $617\pm157$~kpc from their nearest bright neighbour. A difference in local environmental density between the quenched low mass galaxies and star forming comparison is also seen, such that the quenched dwarfs have a local bright ($M_{K} < -23$) galaxy density $\rho_{proj} = 8.2\pm2.0$~Mpc$^{-2}$, compared to $\rho_{proj} = 2.1\pm0.4$~Mpc$^{-2}$ for the star-forming comparison sample. 

This is consistent with previous studies \citep[e.g.][]{2012ApJ...757...85G}, and confirms that low-mass galaxies must be located at a small separation from a massive galaxy (or in over-dense regions of the Universe) for their star formation to be shut-off, with 97~percent of quenched dwarfs in SDSS found within 4~virial radii of a bright/massive host galaxy. While \citet{2012ApJ...757...85G} use a lower mass/luminosity threshold for dwarf galaxies than our work when defining their sample ($<10^{9}$~M$_{\sun}$), this environmental trend holds in the mass range we examine here. 

Due to their low mass, galaxies with $M_{\star} < 5\times10^{9}$~M$_{\sun}$ are likely unable to completely quench their star formation through secular processes (e.g. AGN feedback), and must either undergo interactions with the intra-group/cluster medium, else with a massive galaxy in order for star formation to be permanently quenched. Indeed, only two of the low mass quenched galaxies in our sample, are located more than 1.5~Mpc from the nearest massive galaxy. Internally driven processes such as e.g. maintenance-mode AGN feedback may help these low-mass galaxies maintain quiescence \citep[e.g.][]{2016Natur.533..504C}. 

These two remaining galaxies,  MaNGA 1-322087 and 1-92638 have projected separations $>1.5$~Mpc from the nearest galaxy brighter than $M_{K} = -23$. However, our velocity cut of $\pm1000$~km~s$^{-1}$ when identifying host galaxies is quite restrictive, as galaxies on highly radial plunging or infall orbits can have a large velocity separation from their host galaxies. Relaxing this velocity cut to $\pm1500$~km~s$^{-1}$ does not reveal any closer bright galaxies with $M_{K} < 23$. However, both faint quenched galaxies are within a few virial radii of a nearby galaxy group or cluster, and may have been tidally stripped of star-forming material through a process such as ram pressure stripping. We also cannot rule out the possibility that these galaxies are in a quiescent phase, having ejected their gas due to feedback effects. An analysis of their stellar populations is required to constrain their formation timescales.

We have shown in Fig.~\ref{sizemag} and Fig.~\ref{siglum} that despite being located across multiple groups and clusters, the MaNGA faint quenched  galaxies follow the same size-magnitude and $\sigma$-luminosity relation as bright dE galaxies examined in previous studies.This agrees well with the results of \citet{2015MNRAS.453.3635P}, where dEs in groups and clusters are found to lie on the same regions of the $\sigma$-luminosity and size-magnitude relations.  According to the numerical simulations of \citet{2015MNRAS.454.2502S}, harassment is only efficient in truncating stellar discs in the cluster core environment, and we therefore suggest that group dEs are more likely to host disc-like substructure than their cluster counterparts. Harassment can set up counter-rotating components in a galaxy \citep{2005A&A...444..803G}, though the conditions required for this KDC formation scenario are restrictive.

\subsection{Kinematically distinct cores as evidence of accretion}

We identify two faint galaxies with kinematically distinct cores (KDCs): MaNGA~1-113520 and MaNGA~1-255220. The properties of both objects are described in detail in \ref{sec:kdet}, As can be seen in Fig.~\ref{fig:kindist}, both galaxies have strong peaks in their velocity dispersion profiles, at radii of 1.2~kpc and 1.6~kpc respectively (1.3~$R_{e}$ and 1.7~$R_{e}$ respectively). The $\sigma$ peaks have amplitudes $\sim85$~km~s$^{-1}$ and $\sim110$~km~s$^{-1}$, though $\sigma$ within $1~R_{e}$ is lower than this (67~km~s$^{-1}$ and 83~km~s$^{-1}$ respectively). Both galaxies exhibit strong central $\sigma$ dips of $\sim30$~km~s$^{-1}$, and have rotation $<15$~km~s$^{-1}$ inside 1~$R_{e}$ ($\approx 1$~kpc). Such galaxies are referred to as 2$\sigma$ galaxies in \citet{2011MNRAS.414.2923K}. 

\citet{2011MNRAS.417.1643K} identified a similar low-mass galaxy in the Virgo Cluster, VCC 1475, which also exhibits a strong central dip in its kinematics. Using IFU spectroscopy, \citet{2015ApJ...804...70G} confirm the galaxy's unusual kinematic structure. The galaxies contain two stellar discs, with their rotation offset by $180^{\circ}$, with the $\sigma$ peaks located where the two discs intersect. Indeed, several quenched low-mass galaxies with kinematically distinct cores are now known in the literature \citep{2005A&A...438..491D,2006A&A...445L..19T, 2011MNRAS.417.1643K,2011A&A...526A.114T, 2013MNRAS.428.2980R,2014ApJ...783..120T}. In low density environments, $14\pm10.1$~per~cent of dE galaxies contain KDCs, compared to $5.6\pm2.3$~per~cent of dEs in Virgo \citep{2014ApJ...783..120T}. This is remarkably similar to the fraction of KDCs found in this work, with $5.1 \pm 3.6$~per~cent of our galaxies containing such features. While currently dominated by low-number statistics, it appears KDCs in low mass galaxies  may be more common in low density regions of the Universe versus clusters. 

The formation of such galaxies is suggested to be the result of accretion, either through the accretion of a satellite galaxy, or via the accretion of cold gas. Our nearest bright neighbour search reveals both KDCs to reside in group-like environments, with $<7$ bright neighbour galaxies within 1.5~Mpc, where conditions are likely favourable for such interactions. However, in the cluster environment gas accretion is unlikely, and galaxy mergers are expected to be rare due to the high relative velocities of the cluster members.  We might therefore expect to find more quenched low mass galaxies with counter-rotating stellar components in groups rather than clusters, as conditions are more favourable to their formation via gas accretion. 

A second formation scenario is galaxy harassment, which operates most efficiently in rich galaxy clusters. Harassment produces kinematically decoupled features with a velocity turnover at radii $R > 1.5$~$R_{e}$ \citep{2005A&A...444..803G}, while the KDCs seen in the two objects examined here have a velocity turnover within  $\sim1$~R$_{e}$. Furthermore, the initial conditions required for a low mass galaxy to form a KDC though harassment are very restrictive, and only 1~per~cent of dwarfs in a Virgo-like cluster are expected to exhibit counter-rotating cores \citep{2005A&A...444..803G} if they are the result of galaxy harassment. MaNGA has not yet observed many cluster galaxies, and the galaxies examined here are likely to primarily reside in the group environment. In galaxy groups, conditions are more favourable for gas and/or satellite accretion than in clusters, due to smaller relative velocities between the group galaxies. However, \citet{2011ApJ...726...98K} also show that these kinematically distinct structures can be the result of tidal stirring, whereby repeated interactions with a single, high-mass galaxy such as the Milky Way are able to morphologically transform a low mass disc galaxy. While the tidal stirring simulations of  \citet{2011ApJ...726...98K} focus on much lower mass galaxies than we examine in this work, tidal stirring has been suggested as the origin of early-type dwarfs in groups \citep{2014ApJ...796L..14P}. Thus we cannot completely rule out a tidal origin for the KDCs we identify here.

 \subsection{Non-rotating galaxies}
 
It is clear from the velocity maps in Figure~\ref{fig:kinprofs} and Appendix~\ref{appen1} that the majority of the faint quenched galaxies we examine here are rotating at some level. However, we identify 6 objects that are not rotating at a significant level within 1~$R_{e}$, with $v_{e,rot} < 15$~km~s$^{-1}$: MaNGA~1-38189, MaNGA~1-255220, MaNGA~1-256125, MaNGA~1-258746, MaNGA~1-133948, and MaNGA~1-113520. All have absolute magnitudes $M_{r} \approx -18.5$, and $R_{e}$ between 1~kpc and 1.5~kpc, typical for their luminosity as shown in Fig.~\ref{sizemag}. These galaxies with rotation $<15$~km~s$^{-1}$ at 1~$R_{e}$, have stellar velocity dispersions which vary from $<40$~km~s$^{-1}$ through to 103~km~s$^{-1}$. Most are remarkably round systems, with an axial ratios $b/a > 0.75$ at their effective radii, and thus we may be viewing these dwarfs near face-on, where rotation can be difficult to detect. 
  
Galaxy-galaxy interactions are expected to be extremely common in the core of rich clusters, and thus harassment can easily transform a disc galaxy into a spheroid. This transformation can also act on the galaxy's stellar kinematics, and transform a rotationally supported galaxy into a pressure supported one. For example, in the Virgo Cluster, the majority of the dwarf slow rotators (10/11) examined by \citet{2015ApJ...799..172T} are located in the inner 1~Mpc of the cluster, well within the virial radius, with the same environmental dependence found for the non-dwarf early-type galaxies \citep{2011MNRAS.416.1680C}.  The six faint galaxies we identify here with little/no rotation are all located in groups, where harassment is not thought to be an efficient process in transforming galaxy morphology.  Indeed, they are located at varying distances from their nearest bright neighbour, with projected separations $170~\textrm{kpc} < D_{L{\star}} <1530$~kpc. 

However, harassment is not the only tidal process that can disrupt rotation in a satellite galaxy. The simulations of \citet{2015ApJ...810..100L} show that a low mass disc galaxy in the halo of a Milky-Way like galaxy halo can have its rotation disrupted by tidal stirring. During tidal stirring, bar instabilities are set-up in the disc of the dE progenitor, and this bar eventually collapses into a spheroid. This processes results in dE supported by the random motions of its stars. \citet{2015ApJ...799..171B} suggest that tidal stirring is the cause of the observed environmental trends in dwarf galaxies in Virgo, such that the slow rotators are primarily located in the inner regions of the cluster. Likewise, a similar trend is observed for higher mass galaxies, with slow-rotators nearly absent in the lowest density environments,  and most common in the cluster core \citep{2011MNRAS.416.1680C}.
 The effectiveness of harassment is highly dependent on the nature of the dwarf's orbit, such that a dwarf on a prograde orbit will efficiently remove rotation, while the kinematics will be little affected for a dwarf on a retrograde orbit around its host galaxy.

\subsection{Disc galaxies as the progenitors of faint passive galaxies}

The majority (33/39) of the galaxies in our sample of low mass, quenched galaxies are rotating with $v_{e,rot} > 15$~km~s$^{-1}$, consistent with observations of dE galaxies in the outskirts of the Virgo cluster \citep[e.g.][]{2015ApJ...799..172T}. Rotation in dE galaxies is likely evidence of a disc origin, though interactions with other galaxies and/or the cluster/group tidal potential can erase regular rotation and substructure. The simulations of \citet{2016ApJ...820..131E} show that Internal feedback processes that drive inflow/outflow can drive radial stellar migration, again affecting the galaxy dynamics.

Several of the objects in our faint quenched galaxy sample exhibit clear spiral or disc-like structure. For example, MaNGA-24124 has clear bar/spiral structure at its centre, while MaNGA-113525 has faint spiral arms. \cite{2006AJ....132..497L} show that a high fraction of Virgo dEs exhibit similar structures, despite having red colours and quenched stellar populations. As the MaNGA faint quenched galaxies are typically located in galaxy groups rather than clusters, we might expect more spiral/disc features to survive the transformation process from a star-forming disc to a quenched dwarf. A low mass group dwarf will undergo fewer high-speed tidal interactions than a similar galaxy in a massive cluster core.  

While spiral galaxies are typically assumed to be star forming, spiral galaxies can host optically red stellar populations, consistent with little or no star formation \citep{2009MNRAS.393.1302W,2010MNRAS.405..783M}. \citet{2009MNRAS.393.1324B} show that the fraction of red spiral galaxies with  $M_{\star} \lesssim 5\times10^{9}$~M$_{\sun}$ increases rapidly as local galaxy density increases. At local galaxy densities of $10$~Mpc$^{-2}$, half of all spiral galaxies with $M_{\star} \approx 5\times10^{9}$~M$_{\sun}$ have red optical colours. The fraction of low mass red galaxies that originated as spiral galaxies in these dense environments may well be higher, as spiral structure is easily erased in high density environments. This is in agreement with \citet{2009MNRAS.393.1302W}, who find that below $M_{\star} = 10^{10}$~M$_{\sun}$, red spirals are rare in the Abell 901/902 cluster, with their rapid quenching accompanied by fast morphological transformation. Thus a large fraction of the quenched low-mass galaxies that we observe in groups and clusters today are an extension of this red spiral population to low luminosity. 

\section{Conclusions}
\label{sec:conclude}

We have examined the kinematics of 39 quenched low luminosity galaxies in the MaNGA survey with $M_{r} > -19.1$. The majority (37/39) are located within $\sim1.5$~Mpc and $\pm1000$~km~s$^{-1}$ of a luminous neighbour with $M_{K} > -23$, with a median separation of 184~kpc, confirming the result of \citet{2012ApJ...757...85G} that a massive neighbour is required to quench star formation in low mass galaxies. The faint quenched galaxies have half-light radii $0.64~\textrm{kpc} < R_{e} <  2.57$~kpc, and velocity dispersions ($<$)40~km~s$^{-1} < \sigma_{e} < 128$~km~s$^{-1}$. The majority of our sample  lie on the size-magnitude and $\sigma$-luminosity relation for bright dwarf elliptical galaxies, though five objects with $\sigma > 100$~km~s$^{-1}$ are likely low-luminosity ``classical'' ellipticals rather than dwarf galaxies. 

Kinematic maps reveal the majority of the low-mass quenched MaNGA galaxies to be rotating, with clear disc-like structure seen in many dwarf galaxies of comparable magnitude. Given this result, and that the low mass quenched galaxies are primarily found in galaxy groups/clusters at small separation from a bright neighbour galaxy, we suggest that most of our sample are quenched/passive spiral galaxies that have been stripped of star forming material, but have not undergone sufficient tidal interactions to completely erase their substructure. This is in contrast to smooth dEs observed in clusters such as e.g. Virgo where galaxy harassment is able to operate more efficiently to morphologically transform low-mass disc galaxies into dEs.

Two galaxies, MaNGA~1-113520 and MaNGA~1-255220 show counter-rotation in their stellar kinematics with their velocity profiles  turning over at $<1$~$R_{e}$, and furthermore exhibit distinctive 2$\sigma$ peaks in their stellar velocity dispersion maps.  These two objects host counter-rotating stellar discs, likely the result of a recent gas accretion event, but may have a tidal origin. As we have a relatively small sample size of 39 galaxies, this result suggests low mass galaxies hosting kinematically decoupled cores may be relatively common. 

While the majority of low-mass quenched galaxies in our sample are emission-line free objects, six exhibit weakly-ionised gas emission throughout their structures, with $H\alpha_{EW} < 2$~$\textrm{\AA}$. The two galaxies with KDCs exhibit rotation in their ionised gas component, and this the gas is kinematically misaligned with their stellar component. Four other dwarfs in our sample also have measurable rotation in their ionised gas component, again kinematically-offset from their stellar components. As this gas does not share the same angular momentum as the stellar component, we suggest this kinematic misalignment is the result of accretion. However, we also note that counter-rotating gas that is kinematically offset by $180^{\circ}$ can occur without the need for accretion due to relaxation.

MaNGA has observed 39 quenched low mass galaxies in its first 12 months of observing, we can expect this sample size to grow to $\sim200$ galaxies upon completion of the project. For the first time, the spatially resolved dynamics and star formation histories of low mass galaxies will be examined using a statistically significant sample, and compared to their high-mass counterparts. In future work, for those low-mass galaxies with sufficiently high velocity dispersions, we will determine the fraction of fast vs. slow rotators in this mass regime, and compare these fractions to those found in high-mass galaxies. We will also examine the quenching timescales of the faint galaxy sample. 

\section{Acknowledgements}

We thank the referee for taking time to review this paper, and their comments led to the improvement of this work.

SJP acknowledges postdoctoral funding from the University of Portsmouth. 
AW acknowledges support of a Leverhulme Early Career Fellowship.
This work was supported by World Premier International Research Center Initiative (WPI Initiative), MEXT, Japan.
J.~F-B. acknowledges support from grant AYA2013-48226-C3-1-P from the Spanish Ministry of Economy and Competitiveness (MINECO), as well as from the FP7 Marie Curie Actions of the European Commission, via the Initial Training Network DAGAL under REA grant agreement number 289313.
MAB acknowledges support from NSF AST 1517006.

Funding for the Sloan Digital Sky Survey IV has been provided by
the Alfred P. Sloan Foundation, the U.S. Department of Energy Office of
Science, and the Participating Institutions. SDSS-IV acknowledges
support and resources from the Center for High-Performance Computing at
the University of Utah. The SDSS web site is www.sdss.org.

SDSS-IV is managed by the Astrophysical Research Consortium for the 
Participating Institutions of the SDSS Collaboration including the 
Brazilian Participation Group, the Carnegie Institution for Science, 
Carnegie Mellon University, the Chilean Participation Group, the French Participation Group, Harvard-Smithsonian Center for Astrophysics, 
Instituto de Astrof\'isica de Canarias, The Johns Hopkins University, 
Kavli Institute for the Physics and Mathematics of the Universe (IPMU) / 
University of Tokyo, Lawrence Berkeley National Laboratory, 
Leibniz Institut f\"ur Astrophysik Potsdam (AIP),  
Max-Planck-Institut f\"ur Astronomie (MPIA Heidelberg), 
Max-Planck-Institut f\"ur Astrophysik (MPA Garching), 
Max-Planck-Institut f\"ur Extraterrestrische Physik (MPE), 
National Astronomical Observatory of China, New Mexico State University, 
New York University, University of Notre Dame, 
Observat\'ario Nacional / MCTI, The Ohio State University, 
Pennsylvania State University, Shanghai Astronomical Observatory, 
United Kingdom Participation Group,
Universidad Nacional Aut\'onoma de M\'exico, University of Arizona, 
University of Colorado Boulder, University of Oxford, University of Portsmouth, 
University of Utah, University of Virginia, University of Washington, University of Wisconsin, 
Vanderbilt University, and Yale University.

This research has made use of the NASA/IPAC Extragalactic Database (NED),
which is operated by the Jet Propulsion Laboratory, California Institute of Technology,
under contract with the National Aeronautics and Space Administration.




\bibliographystyle{mnras}
\bibliography{spenny} 

\begin{thebibliography}{}
\makeatletter
\relax
\def\mn@urlcharsother{\let\do\@makeother \do\$\do\&\do\#\do\^\do\_\do\%\do\~}
\def\mn@doi{\begingroup\mn@urlcharsother \@ifnextchar [ {\mn@doi@}
  {\mn@doi@[]}}
\def\mn@doi@[#1]#2{\def\@tempa{#1}\ifx\@tempa\@empty \href
  {http://dx.doi.org/#2} {doi:#2}\else \href {http://dx.doi.org/#2} {#1}\fi
  \endgroup}
\def\mn@eprint#1#2{\mn@eprint@#1:#2::\@nil}
\def\mn@eprint@arXiv#1{\href {http://arxiv.org/abs/#1} {{\tt arXiv:#1}}}
\def\mn@eprint@dblp#1{\href {http://dblp.uni-trier.de/rec/bibtex/#1.xml}
  {dblp:#1}}
\def\mn@eprint@#1:#2:#3:#4\@nil{\def\@tempa {#1}\def\@tempb {#2}\def\@tempc
  {#3}\ifx \@tempc \@empty \let \@tempc \@tempb \let \@tempb \@tempa \fi \ifx
  \@tempb \@empty \def\@tempb {arXiv}\fi \@ifundefined
  {mn@eprint@\@tempb}{\@tempb:\@tempc}{\expandafter \expandafter \csname
  mn@eprint@\@tempb\endcsname \expandafter{\@tempc}}}

\bibitem[\protect\citeauthoryear{{Abadi}, {Moore}  \& {Bower}}{{Abadi}
  et~al.}{1999}]{1999MNRAS.308..947A}
{Abadi} M.~G.,  {Moore} B.,   {Bower} R.~G.,  1999, \mn@doi [\mnras]
  {10.1046/j.1365-8711.1999.02715.x}, \href
  {http://adsabs.harvard.edu/abs/1999MNRAS.308..947A} {308, 947}

\bibitem[\protect\citeauthoryear{{Baldry}, {Balogh}, {Bower}, {Glazebrook},
  {Nichol}, {Bamford}  \& {Budavari}}{{Baldry}
  et~al.}{2006}]{2006MNRAS.373..469B}
{Baldry} I.~K.,  {Balogh} M.~L.,  {Bower} R.~G.,  {Glazebrook} K.,  {Nichol}
  R.~C.,  {Bamford} S.~P.,   {Budavari} T.,  2006, \mn@doi [\mnras]
  {10.1111/j.1365-2966.2006.11081.x}, \href
  {http://adsabs.harvard.edu/abs/2006MNRAS.373..469B} {373, 469}

\bibitem[\protect\citeauthoryear{{Balogh}, {Schade}, {Morris}, {Yee},
  {Carlberg}  \& {Ellingson}}{{Balogh} et~al.}{1998}]{1998ApJ...504L..75B}
{Balogh} M.~L.,  {Schade} D.,  {Morris} S.~L.,  {Yee} H.~K.~C.,  {Carlberg}
  R.~G.,   {Ellingson} E.,  1998, \mn@doi [\apjl] {10.1086/311576}, \href
  {http://adsabs.harvard.edu/abs/1998ApJ...504L..75B} {504, L75}

\bibitem[\protect\citeauthoryear{{Bamford} et~al.,}{{Bamford}
  et~al.}{2009}]{2009MNRAS.393.1324B}
{Bamford} S.~P.,  et~al., 2009, \mn@doi [\mnras]
  {10.1111/j.1365-2966.2008.14252.x}, \href
  {http://adsabs.harvard.edu/abs/2009MNRAS.393.1324B} {393, 1324}

\bibitem[\protect\citeauthoryear{{Belfiore} et~al.,}{{Belfiore}
  et~al.}{2015}]{2015MNRAS.449..867B}
{Belfiore} F.,  et~al., 2015, \mn@doi [\mnras] {10.1093/mnras/stv296}, \href
  {http://adsabs.harvard.edu/abs/2015MNRAS.449..867B} {449, 867}

\bibitem[\protect\citeauthoryear{{Bender}}{{Bender}}{1988}]{1988A&A...202L...5B}
{Bender} R.,  1988, \aap, \href
  {http://adsabs.harvard.edu/abs/1988A%26A...202L...5B} {202, L5}

\bibitem[\protect\citeauthoryear{{Bender} \& {Nieto}}{{Bender} \&
  {Nieto}}{1990}]{1990A&A...239...97B}
{Bender} R.,  {Nieto} J.-L.,  1990, \aap, \href
  {http://adsabs.harvard.edu/abs/1990A%26A...239...97B} {239, 97}

\bibitem[\protect\citeauthoryear{{Benson}, {Toloba}, {Mayer}, {Simon}  \&
  {Guhathakurta}}{{Benson} et~al.}{2015}]{2015ApJ...799..171B}
{Benson} A.~J.,  {Toloba} E.,  {Mayer} L.,  {Simon} J.~D.,   {Guhathakurta} P.,
   2015, \mn@doi [\apj] {10.1088/0004-637X/799/2/171}, \href
  {http://adsabs.harvard.edu/abs/2015ApJ...799..171B} {799, 171}

\bibitem[\protect\citeauthoryear{{Binggeli}, {Tarenghi}  \&
  {Sandage}}{{Binggeli} et~al.}{1990}]{1990A&A...228...42B}
{Binggeli} B.,  {Tarenghi} M.,   {Sandage} A.,  1990, \aap, \href
  {http://adsabs.harvard.edu/abs/1990A%26A...228...42B} {228, 42}

\bibitem[\protect\citeauthoryear{{Blanton} \& {Roweis}}{{Blanton} \&
  {Roweis}}{2007}]{2007AJ....133..734B}
{Blanton} M.~R.,  {Roweis} S.,  2007, \mn@doi [\aj] {10.1086/510127}, \href
  {http://adsabs.harvard.edu/abs/2007AJ....133..734B} {133, 734}

\bibitem[\protect\citeauthoryear{{Blanton}, {Kazin}, {Muna}, {Weaver}  \&
  {Price-Whelan}}{{Blanton} et~al.}{2011}]{2011AJ....142...31B}
{Blanton} M.~R.,  {Kazin} E.,  {Muna} D.,  {Weaver} B.~A.,   {Price-Whelan} A.,
   2011, \mn@doi [\aj] {10.1088/0004-6256/142/1/31}, \href
  {http://adsabs.harvard.edu/abs/2011AJ....142...31B} {142, 31}

\bibitem[\protect\citeauthoryear{{Bundy} et~al.,}{{Bundy}
  et~al.}{2015}]{2015ApJ...798....7B}
{Bundy} K.,  et~al., 2015, \mn@doi [\apj] {10.1088/0004-637X/798/1/7}, \href
  {http://adsabs.harvard.edu/abs/2015ApJ...798....7B} {798, 7}

\bibitem[\protect\citeauthoryear{{Cappellari} \& {Emsellem}}{{Cappellari} \&
  {Emsellem}}{2004}]{2004PASP..116..138C}
{Cappellari} M.,  {Emsellem} E.,  2004, \mn@doi [\pasp] {10.1086/381875}, \href
  {http://adsabs.harvard.edu/abs/2004PASP..116..138C} {116, 138}

\bibitem[\protect\citeauthoryear{{Cappellari} et~al.,}{{Cappellari}
  et~al.}{2011a}]{2011MNRAS.413..813C}
{Cappellari} M.,  et~al., 2011a, \mn@doi [\mnras]
  {10.1111/j.1365-2966.2010.18174.x}, \href
  {http://adsabs.harvard.edu/abs/2011MNRAS.413..813C} {413, 813}

\bibitem[\protect\citeauthoryear{{Cappellari} et~al.,}{{Cappellari}
  et~al.}{2011b}]{2011MNRAS.416.1680C}
{Cappellari} M.,  et~al., 2011b, \mn@doi [\mnras]
  {10.1111/j.1365-2966.2011.18600.x}, \href
  {http://adsabs.harvard.edu/abs/2011MNRAS.416.1680C} {416, 1680}

\bibitem[\protect\citeauthoryear{{Carollo} et~al.,}{{Carollo}
  et~al.}{2016}]{2016ApJ...818..180C}
{Carollo} C.~M.,  et~al., 2016, \mn@doi [\apj] {10.3847/0004-637X/818/2/180},
  \href {http://adsabs.harvard.edu/abs/2016ApJ...818..180C} {818, 180}

\bibitem[\protect\citeauthoryear{{Chabrier}}{{Chabrier}}{2003}]{2003PASP..115..763C}
{Chabrier} G.,  2003, \mn@doi [\pasp] {10.1086/376392}, \href
  {http://adsabs.harvard.edu/abs/2003PASP..115..763C} {115, 763}

\bibitem[\protect\citeauthoryear{{Cheung} et~al.,}{{Cheung}
  et~al.}{2016}]{2016Natur.533..504C}
{Cheung} E.,  et~al., 2016, \mn@doi [\nat] {10.1038/nature18006}, \href
  {http://adsabs.harvard.edu/abs/2016Natur.533..504C} {533, 504}

\bibitem[\protect\citeauthoryear{{Conselice}, {Gallagher}  \&
  {Wyse}}{{Conselice} et~al.}{2001}]{2001ApJ...559..791C}
{Conselice} C.~J.,  {Gallagher} III J.~S.,   {Wyse} R.~F.~G.,  2001, \mn@doi
  [\apj] {10.1086/322373}, \href
  {http://adsabs.harvard.edu/abs/2001ApJ...559..791C} {559, 791}

\bibitem[\protect\citeauthoryear{{Croton} \& {Farrar}}{{Croton} \&
  {Farrar}}{2008}]{2008MNRAS.386.2285C}
{Croton} D.~J.,  {Farrar} G.~R.,  2008, \mn@doi [\mnras]
  {10.1111/j.1365-2966.2008.13204.x}, \href
  {http://adsabs.harvard.edu/abs/2008MNRAS.386.2285C} {386, 2285}

\bibitem[\protect\citeauthoryear{{Croton} et~al.,}{{Croton}
  et~al.}{2006}]{2006MNRAS.365...11C}
{Croton} D.~J.,  et~al., 2006, \mn@doi [\mnras]
  {10.1111/j.1365-2966.2005.09675.x}, \href
  {http://adsabs.harvard.edu/abs/2006MNRAS.365...11C} {365, 11}

\bibitem[\protect\citeauthoryear{{Darvish}, {Mobasher}, {Sobral}, {Rettura},
  {Scoville}, {Faisst}  \& {Capak}}{{Darvish}
  et~al.}{2016}]{2016arXiv160503182D}
{Darvish} B.,  {Mobasher} B.,  {Sobral} D.,  {Rettura} A.,  {Scoville} N.,
  {Faisst} A.,   {Capak} P.,  2016, preprint, \href
  {http://adsabs.harvard.edu/abs/2016arXiv160503182D} {} (\mn@eprint {arXiv}
  {1605.03182})

\bibitem[\protect\citeauthoryear{{Davidzon} et~al.,}{{Davidzon}
  et~al.}{2016}]{2016A&A...586A..23D}
{Davidzon} I.,  et~al., 2016, \mn@doi [\aap] {10.1051/0004-6361/201527129},
  \href {http://adsabs.harvard.edu/abs/2016A%26A...586A..23D} {586, A23}

\bibitem[\protect\citeauthoryear{{Davies}, {Efstathiou}, {Fall}, {Illingworth}
  \& {Schechter}}{{Davies} et~al.}{1983}]{1983ApJ...266...41D}
{Davies} R.~L.,  {Efstathiou} G.,  {Fall} S.~M.,  {Illingworth} G.,
  {Schechter} P.~L.,  1983, \mn@doi [\apj] {10.1086/160757}, \href
  {http://adsabs.harvard.edu/abs/1983ApJ...266...41D} {266, 41}

\bibitem[\protect\citeauthoryear{{Davis} et~al.,}{{Davis}
  et~al.}{2011}]{2011MNRAS.417..882D}
{Davis} T.~A.,  et~al., 2011, \mn@doi [\mnras]
  {10.1111/j.1365-2966.2011.19355.x}, \href
  {http://adsabs.harvard.edu/abs/2011MNRAS.417..882D} {417, 882}

\bibitem[\protect\citeauthoryear{{De Rijcke}, {Dejonghe}, {Zeilinger}  \&
  {Hau}}{{De Rijcke} et~al.}{2003}]{2003A&A...400..119D}
{De Rijcke} S.,  {Dejonghe} H.,  {Zeilinger} W.~W.,   {Hau} G.~K.~T.,  2003,
  \mn@doi [\aap] {10.1051/0004-6361:20021866}, \href
  {http://adsabs.harvard.edu/abs/2003A%26A...400..119D} {400, 119}

\bibitem[\protect\citeauthoryear{{Dekel} \& {Silk}}{{Dekel} \&
  {Silk}}{1986}]{1986ApJ...303...39D}
{Dekel} A.,  {Silk} J.,  1986, \mn@doi [\apj] {10.1086/164050}, \href
  {http://adsabs.harvard.edu/abs/1986ApJ...303...39D} {303, 39}

\bibitem[\protect\citeauthoryear{{Drinkwater}, {Gregg}  \&
  {Colless}}{{Drinkwater} et~al.}{2001}]{2001ApJ...548L.139D}
{Drinkwater} M.~J.,  {Gregg} M.~D.,   {Colless} M.,  2001, \mn@doi [\apjl]
  {10.1086/319113}, \href {http://adsabs.harvard.edu/abs/2001ApJ...548L.139D}
  {548, L139}

\bibitem[\protect\citeauthoryear{{Drory} et~al.,}{{Drory}
  et~al.}{2015}]{2015AJ....149...77D}
{Drory} N.,  et~al., 2015, \mn@doi [\aj] {10.1088/0004-6256/149/2/77}, \href
  {http://adsabs.harvard.edu/abs/2015AJ....149...77D} {149, 77}

\bibitem[\protect\citeauthoryear{{Eggen}, {Lynden-Bell}  \& {Sandage}}{{Eggen}
  et~al.}{1962}]{1962ApJ...136..748E}
{Eggen} O.~J.,  {Lynden-Bell} D.,   {Sandage} A.~R.,  1962, \mn@doi [\apj]
  {10.1086/147433}, \href {http://adsabs.harvard.edu/abs/1962ApJ...136..748E}
  {136, 748}

\bibitem[\protect\citeauthoryear{{El-Badry}, {Wetzel}, {Geha}, {Hopkins},
  {Kere{\v s}}, {Chan}  \& {Faucher-Gigu{\`e}re}}{{El-Badry}
  et~al.}{2016}]{2016ApJ...820..131E}
{El-Badry} K.,  {Wetzel} A.,  {Geha} M.,  {Hopkins} P.~F.,  {Kere{\v s}} D.,
  {Chan} T.~K.,   {Faucher-Gigu{\`e}re} C.-A.,  2016, \mn@doi [\apj]
  {10.3847/0004-637X/820/2/131}, \href
  {http://adsabs.harvard.edu/abs/2016ApJ...820..131E} {820, 131}

\bibitem[\protect\citeauthoryear{{Emsellem} et~al.,}{{Emsellem}
  et~al.}{2011}]{2011MNRAS.414..888E}
{Emsellem} E.,  et~al., 2011, \mn@doi [\mnras]
  {10.1111/j.1365-2966.2011.18496.x}, \href
  {http://adsabs.harvard.edu/abs/2011MNRAS.414..888E} {414, 888}

\bibitem[\protect\citeauthoryear{{Forbes}, {Spitler}, {Graham}, {Foster}, {Hau}
   \& {Benson}}{{Forbes} et~al.}{2011}]{2011MNRAS.413.2665F}
{Forbes} D.~A.,  {Spitler} L.~R.,  {Graham} A.~W.,  {Foster} C.,  {Hau}
  G.~K.~T.,   {Benson} A.,  2011, \mn@doi [\mnras]
  {10.1111/j.1365-2966.2011.18335.x}, \href
  {http://adsabs.harvard.edu/abs/2011MNRAS.413.2665F} {413, 2665}

\bibitem[\protect\citeauthoryear{{Geha}, {Guhathakurta}  \& {van der
  Marel}}{{Geha} et~al.}{2003}]{2003AJ....126.1794G}
{Geha} M.,  {Guhathakurta} P.,   {van der Marel} R.~P.,  2003, \mn@doi [\aj]
  {10.1086/377624}, \href {http://adsabs.harvard.edu/abs/2003AJ....126.1794G}
  {126, 1794}

\bibitem[\protect\citeauthoryear{{Geha}, {Blanton}, {Yan}  \& {Tinker}}{{Geha}
  et~al.}{2012}]{2012ApJ...757...85G}
{Geha} M.,  {Blanton} M.~R.,  {Yan} R.,   {Tinker} J.~L.,  2012, \mn@doi [\apj]
  {10.1088/0004-637X/757/1/85}, \href
  {http://adsabs.harvard.edu/abs/2012ApJ...757...85G} {757, 85}

\bibitem[\protect\citeauthoryear{{Gonz{\'a}lez-Garc{\'{\i}}a}, {Aguerri}  \&
  {Balcells}}{{Gonz{\'a}lez-Garc{\'{\i}}a} et~al.}{2005}]{2005A&A...444..803G}
{Gonz{\'a}lez-Garc{\'{\i}}a} A.~C.,  {Aguerri} J.~A.~L.,   {Balcells} M.,
  2005, \mn@doi [\aap] {10.1051/0004-6361:20052670}, \href
  {http://adsabs.harvard.edu/abs/2005A%26A...444..803G} {444, 803}

\bibitem[\protect\citeauthoryear{{Gonz{\'a}lez-Samaniego}, {Col{\'{\i}}n},
  {Avila-Reese}, {Rodr{\'{\i}}guez-Puebla}  \&
  {Valenzuela}}{{Gonz{\'a}lez-Samaniego} et~al.}{2014}]{2014ApJ...785...58G}
{Gonz{\'a}lez-Samaniego} A.,  {Col{\'{\i}}n} P.,  {Avila-Reese} V.,
  {Rodr{\'{\i}}guez-Puebla} A.,   {Valenzuela} O.,  2014, \mn@doi [\apj]
  {10.1088/0004-637X/785/1/58}, \href
  {http://adsabs.harvard.edu/abs/2014ApJ...785...58G} {785, 58}

\bibitem[\protect\citeauthoryear{{Graham}, {Jerjen}  \& {Guzm{\'a}n}}{{Graham}
  et~al.}{2003}]{2003AJ....126.1787G}
{Graham} A.~W.,  {Jerjen} H.,   {Guzm{\'a}n} R.,  2003, \mn@doi [\aj]
  {10.1086/378166}, \href {http://adsabs.harvard.edu/abs/2003AJ....126.1787G}
  {126, 1787}

\bibitem[\protect\citeauthoryear{{Gu{\'e}rou} et~al.,}{{Gu{\'e}rou}
  et~al.}{2015}]{2015ApJ...804...70G}
{Gu{\'e}rou} A.,  et~al., 2015, \mn@doi [\apj] {10.1088/0004-637X/804/1/70},
  \href {http://adsabs.harvard.edu/abs/2015ApJ...804...70G} {804, 70}

\bibitem[\protect\citeauthoryear{{Gunn} \& {Gott}}{{Gunn} \&
  {Gott}}{1972}]{1972ApJ...176....1G}
{Gunn} J.~E.,  {Gott} III J.~R.,  1972, \mn@doi [\apj] {10.1086/151605}, \href
  {http://adsabs.harvard.edu/abs/1972ApJ...176....1G} {176, 1}

\bibitem[\protect\citeauthoryear{{Gunn} et~al.,}{{Gunn}
  et~al.}{2006}]{2006AJ....131.2332G}
{Gunn} J.~E.,  et~al., 2006, \mn@doi [\aj] {10.1086/500975}, \href
  {http://adsabs.harvard.edu/abs/2006AJ....131.2332G} {131, 2332}

\bibitem[\protect\citeauthoryear{{Huchra} et~al.,}{{Huchra}
  et~al.}{2012}]{2012ApJS..199...26H}
{Huchra} J.~P.,  et~al., 2012, \mn@doi [\apjs] {10.1088/0067-0049/199/2/26},
  \href {http://adsabs.harvard.edu/abs/2012ApJS..199...26H} {199, 26}

\bibitem[\protect\citeauthoryear{{Janz} et~al.,}{{Janz}
  et~al.}{2012}]{2012ApJ...745L..24J}
{Janz} J.,  et~al., 2012, \mn@doi [\apjl] {10.1088/2041-8205/745/2/L24}, \href
  {http://adsabs.harvard.edu/abs/2012ApJ...745L..24J} {745, L24}

\bibitem[\protect\citeauthoryear{{Kauffmann} et~al.,}{{Kauffmann}
  et~al.}{2003}]{2003MNRAS.341...54K}
{Kauffmann} G.,  et~al., 2003, \mn@doi [\mnras]
  {10.1046/j.1365-8711.2003.06292.x}, \href
  {http://adsabs.harvard.edu/abs/2003MNRAS.341...54K} {341, 54}

\bibitem[\protect\citeauthoryear{{Kazantzidis}, {{\L}okas}, {Callegari},
  {Mayer}  \& {Moustakas}}{{Kazantzidis} et~al.}{2011}]{2011ApJ...726...98K}
{Kazantzidis} S.,  {{\L}okas} E.~L.,  {Callegari} S.,  {Mayer} L.,
  {Moustakas} L.~A.,  2011, \mn@doi [\apj] {10.1088/0004-637X/726/2/98}, \href
  {http://adsabs.harvard.edu/abs/2011ApJ...726...98K} {726, 98}

\bibitem[\protect\citeauthoryear{{Koleva}, {Prugniel}, {de Rijcke}  \&
  {Zeilinger}}{{Koleva} et~al.}{2011}]{2011MNRAS.417.1643K}
{Koleva} M.,  {Prugniel} P.,  {de Rijcke} S.,   {Zeilinger} W.~W.,  2011,
  \mn@doi [\mnras] {10.1111/j.1365-2966.2011.19057.x}, \href
  {http://adsabs.harvard.edu/abs/2011MNRAS.417.1643K} {417, 1643}

\bibitem[\protect\citeauthoryear{{Kormendy} \& {Bender}}{{Kormendy} \&
  {Bender}}{2012}]{2012ApJS..198....2K}
{Kormendy} J.,  {Bender} R.,  2012, \mn@doi [\apjs]
  {10.1088/0067-0049/198/1/2}, \href
  {http://adsabs.harvard.edu/abs/2012ApJS..198....2K} {198, 2}

\bibitem[\protect\citeauthoryear{{Kormendy} \& {Illingworth}}{{Kormendy} \&
  {Illingworth}}{1982}]{1982ApJ...256..460K}
{Kormendy} J.,  {Illingworth} G.,  1982, \mn@doi [\apj] {10.1086/159923}, \href
  {http://adsabs.harvard.edu/abs/1982ApJ...256..460K} {256, 460}

\bibitem[\protect\citeauthoryear{{Krajnovi{\'c}} et~al.,}{{Krajnovi{\'c}}
  et~al.}{2011}]{2011MNRAS.414.2923K}
{Krajnovi{\'c}} D.,  et~al., 2011, \mn@doi [\mnras]
  {10.1111/j.1365-2966.2011.18560.x}, \href
  {http://adsabs.harvard.edu/abs/2011MNRAS.414.2923K} {414, 2923}

\bibitem[\protect\citeauthoryear{{Larson}, {Tinsley}  \& {Caldwell}}{{Larson}
  et~al.}{1980}]{1980ApJ...237..692L}
{Larson} R.~B.,  {Tinsley} B.~M.,   {Caldwell} C.~N.,  1980, \mn@doi [\apj]
  {10.1086/157917}, \href {http://adsabs.harvard.edu/abs/1980ApJ...237..692L}
  {237, 692}

\bibitem[\protect\citeauthoryear{{Law} et~al.,}{{Law}
  et~al.}{2015}]{2015AJ....150...19L}
{Law} D.~R.,  et~al., 2015, \mn@doi [\aj] {10.1088/0004-6256/150/1/19}, \href
  {http://adsabs.harvard.edu/abs/2015AJ....150...19L} {150, 19}

\bibitem[\protect\citeauthoryear{{Lewis} et~al.,}{{Lewis}
  et~al.}{2002}]{2002MNRAS.334..673L}
{Lewis} I.,  et~al., 2002, \mn@doi [\mnras] {10.1046/j.1365-8711.2002.05558.x},
  \href {http://adsabs.harvard.edu/abs/2002MNRAS.334..673L} {334, 673}

\bibitem[\protect\citeauthoryear{{Li} et~al.,}{{Li}
  et~al.}{2015}]{2015ApJ...804..125L}
{Li} C.,  et~al., 2015, \mn@doi [\apj] {10.1088/0004-637X/804/2/125}, \href
  {http://adsabs.harvard.edu/abs/2015ApJ...804..125L} {804, 125}

\bibitem[\protect\citeauthoryear{{Lisker}, {Grebel}  \& {Binggeli}}{{Lisker}
  et~al.}{2006}]{2006AJ....132..497L}
{Lisker} T.,  {Grebel} E.~K.,   {Binggeli} B.,  2006, \mn@doi [\aj]
  {10.1086/505045}, \href {http://adsabs.harvard.edu/abs/2006AJ....132..497L}
  {132, 497}

\bibitem[\protect\citeauthoryear{{Lisker} et~al.,}{{Lisker}
  et~al.}{2009}]{2009ApJ...706L.124L}
{Lisker} T.,  et~al., 2009, \mn@doi [\apjl] {10.1088/0004-637X/706/1/L124},
  \href {http://adsabs.harvard.edu/abs/2009ApJ...706L.124L} {706, L124}

\bibitem[\protect\citeauthoryear{{{\L}okas}, {Semczuk}, {Gajda}  \&
  {D'Onghia}}{{{\L}okas} et~al.}{2015}]{2015ApJ...810..100L}
{{\L}okas} E.~L.,  {Semczuk} M.,  {Gajda} G.,   {D'Onghia} E.,  2015, \mn@doi
  [\apj] {10.1088/0004-637X/810/2/100}, \href
  {http://adsabs.harvard.edu/abs/2015ApJ...810..100L} {810, 100}

\bibitem[\protect\citeauthoryear{{Madau}, {Ferguson}, {Dickinson},
  {Giavalisco}, {Steidel}  \& {Fruchter}}{{Madau}
  et~al.}{1996}]{1996MNRAS.283.1388M}
{Madau} P.,  {Ferguson} H.~C.,  {Dickinson} M.~E.,  {Giavalisco} M.,  {Steidel}
  C.~C.,   {Fruchter} A.,  1996, \mn@doi [\mnras] {10.1093/mnras/283.4.1388},
  \href {http://adsabs.harvard.edu/abs/1996MNRAS.283.1388M} {283, 1388}

\bibitem[\protect\citeauthoryear{{Maraston} \& {Str{\"o}mb{\"a}ck}}{{Maraston}
  \& {Str{\"o}mb{\"a}ck}}{2011}]{2011MNRAS.418.2785M}
{Maraston} C.,  {Str{\"o}mb{\"a}ck} G.,  2011, \mn@doi [\mnras]
  {10.1111/j.1365-2966.2011.19738.x}, \href
  {http://adsabs.harvard.edu/abs/2011MNRAS.418.2785M} {418, 2785}

\bibitem[\protect\citeauthoryear{{Martin} et~al.,}{{Martin}
  et~al.}{2005}]{2005ApJ...619L...1M}
{Martin} D.~C.,  et~al., 2005, \mn@doi [\apjl] {10.1086/426387}, \href
  {http://adsabs.harvard.edu/abs/2005ApJ...619L...1M} {619, L1}

\bibitem[\protect\citeauthoryear{{Masters} et~al.,}{{Masters}
  et~al.}{2010}]{2010MNRAS.405..783M}
{Masters} K.~L.,  et~al., 2010, \mn@doi [\mnras]
  {10.1111/j.1365-2966.2010.16503.x}, \href
  {http://adsabs.harvard.edu/abs/2010MNRAS.405..783M} {405, 783}

\bibitem[\protect\citeauthoryear{{McDermid} et~al.,}{{McDermid}
  et~al.}{2006}]{2006MNRAS.373..906M}
{McDermid} R.~M.,  et~al., 2006, \mn@doi [\mnras]
  {10.1111/j.1365-2966.2006.11065.x}, \href
  {http://adsabs.harvard.edu/abs/2006MNRAS.373..906M} {373, 906}

\bibitem[\protect\citeauthoryear{{Moore}, {Katz}, {Lake}, {Dressler}  \&
  {Oemler}}{{Moore} et~al.}{1996}]{1996Natur.379..613M}
{Moore} B.,  {Katz} N.,  {Lake} G.,  {Dressler} A.,   {Oemler} A.,  1996,
  \mn@doi [\nat] {10.1038/379613a0}, \href
  {http://adsabs.harvard.edu/abs/1996Natur.379..613M} {379, 613}

\bibitem[\protect\citeauthoryear{{Muldrew} et~al.,}{{Muldrew}
  et~al.}{2012}]{2012MNRAS.419.2670M}
{Muldrew} S.~I.,  et~al., 2012, \mn@doi [\mnras]
  {10.1111/j.1365-2966.2011.19922.x}, \href
  {http://adsabs.harvard.edu/abs/2012MNRAS.419.2670M} {419, 2670}

\bibitem[\protect\citeauthoryear{{Muratov}, {Kere{\v s}},
  {Faucher-Gigu{\`e}re}, {Hopkins}, {Quataert}  \& {Murray}}{{Muratov}
  et~al.}{2015}]{2015MNRAS.454.2691M}
{Muratov} A.~L.,  {Kere{\v s}} D.,  {Faucher-Gigu{\`e}re} C.-A.,  {Hopkins}
  P.~F.,  {Quataert} E.,   {Murray} N.,  2015, \mn@doi [\mnras]
  {10.1093/mnras/stv2126}, \href
  {http://adsabs.harvard.edu/abs/2015MNRAS.454.2691M} {454, 2691}

\bibitem[\protect\citeauthoryear{{Naab} et~al.,}{{Naab}
  et~al.}{2014}]{2014MNRAS.444.3357N}
{Naab} T.,  et~al., 2014, \mn@doi [\mnras] {10.1093/mnras/stt1919}, \href
  {http://adsabs.harvard.edu/abs/2014MNRAS.444.3357N} {444, 3357}

\bibitem[\protect\citeauthoryear{{Norris} et~al.,}{{Norris}
  et~al.}{2014}]{2014MNRAS.443.1151N}
{Norris} M.~A.,  et~al., 2014, \mn@doi [\mnras] {10.1093/mnras/stu1186}, \href
  {http://adsabs.harvard.edu/abs/2014MNRAS.443.1151N} {443, 1151}

\bibitem[\protect\citeauthoryear{{Paudel} \& {Ree}}{{Paudel} \&
  {Ree}}{2014}]{2014ApJ...796L..14P}
{Paudel} S.,  {Ree} C.~H.,  2014, \mn@doi [\apjl]
  {10.1088/2041-8205/796/1/L14}, \href
  {http://adsabs.harvard.edu/abs/2014ApJ...796L..14P} {796, L14}

\bibitem[\protect\citeauthoryear{{Paudel}, {Lisker}, {Kuntschner}, {Grebel}  \&
  {Glatt}}{{Paudel} et~al.}{2010}]{2010MNRAS.405..800P}
{Paudel} S.,  {Lisker} T.,  {Kuntschner} H.,  {Grebel} E.~K.,   {Glatt} K.,
  2010, \mn@doi [\mnras] {10.1111/j.1365-2966.2010.16507.x}, \href
  {http://adsabs.harvard.edu/abs/2010MNRAS.405..800P} {405, 800}

\bibitem[\protect\citeauthoryear{{Peng} et~al.,}{{Peng}
  et~al.}{2010}]{2010ApJ...721..193P}
{Peng} Y.-j.,  et~al., 2010, \mn@doi [\apj] {10.1088/0004-637X/721/1/193},
  \href {http://adsabs.harvard.edu/abs/2010ApJ...721..193P} {721, 193}

\bibitem[\protect\citeauthoryear{{Penny} \& {Conselice}}{{Penny} \&
  {Conselice}}{2008}]{2008MNRAS.383..247P}
{Penny} S.~J.,  {Conselice} C.~J.,  2008, \mn@doi [\mnras]
  {10.1111/j.1365-2966.2007.12535.x}, \href
  {http://adsabs.harvard.edu/abs/2008MNRAS.383..247P} {383, 247}

\bibitem[\protect\citeauthoryear{{Penny}, {Forbes}, {Pimbblet}  \&
  {Floyd}}{{Penny} et~al.}{2014}]{2014MNRAS.443.3381P}
{Penny} S.~J.,  {Forbes} D.~A.,  {Pimbblet} K.~A.,   {Floyd} D.~J.~E.,  2014,
  \mn@doi [\mnras] {10.1093/mnras/stu1397}, \href
  {http://adsabs.harvard.edu/abs/2014MNRAS.443.3381P} {443, 3381}

\bibitem[\protect\citeauthoryear{{Penny} et~al.,}{{Penny}
  et~al.}{2015a}]{2015MNRAS.453.3519P}
{Penny} S.~J.,  et~al., 2015a, \mn@doi [\mnras] {10.1093/mnras/stv1926}, \href
  {http://adsabs.harvard.edu/abs/2015MNRAS.453.3519P} {453, 3519}

\bibitem[\protect\citeauthoryear{{Penny}, {Janz}, {Forbes}, {Benson}  \&
  {Mould}}{{Penny} et~al.}{2015b}]{2015MNRAS.453.3635P}
{Penny} S.~J.,  {Janz} J.,  {Forbes} D.~A.,  {Benson} A.~J.,   {Mould} J.,
  2015b, \mn@doi [\mnras] {10.1093/mnras/stv1850}, \href
  {http://adsabs.harvard.edu/abs/2015MNRAS.453.3635P} {453, 3635}

\bibitem[\protect\citeauthoryear{{Quilis}, {Moore}  \& {Bower}}{{Quilis}
  et~al.}{2000}]{2000Sci...288.1617Q}
{Quilis} V.,  {Moore} B.,   {Bower} R.,  2000, \mn@doi [Science]
  {10.1126/science.288.5471.1617}, \href
  {http://adsabs.harvard.edu/abs/2000Sci...288.1617Q} {288, 1617}

\bibitem[\protect\citeauthoryear{{Ry{\'s}}, {Falc{\'o}n-Barroso}  \& {van de
  Ven}}{{Ry{\'s}} et~al.}{2013}]{2013MNRAS.428.2980R}
{Ry{\'s}} A.,  {Falc{\'o}n-Barroso} J.,   {van de Ven} G.,  2013, \mn@doi
  [\mnras] {10.1093/mnras/sts245}, \href
  {http://adsabs.harvard.edu/abs/2013MNRAS.428.2980R} {428, 2980}

\bibitem[\protect\citeauthoryear{{Ry{\'s}}, {Koleva}, {Falc{\'o}n-Barroso},
  {Vazdekis}, {Lisker}, {Peletier}  \& {van de Ven}}{{Ry{\'s}}
  et~al.}{2015}]{2015MNRAS.452.1888R}
{Ry{\'s}} A.,  {Koleva} M.,  {Falc{\'o}n-Barroso} J.,  {Vazdekis} A.,  {Lisker}
  T.,  {Peletier} R.,   {van de Ven} G.,  2015, \mn@doi [\mnras]
  {10.1093/mnras/stv1364}, \href
  {http://adsabs.harvard.edu/abs/2015MNRAS.452.1888R} {452, 1888}

\bibitem[\protect\citeauthoryear{{S{\'a}nchez-Bl{\'a}zquez}
  et~al.,}{{S{\'a}nchez-Bl{\'a}zquez} et~al.}{2006}]{2006MNRAS.371..703S}
{S{\'a}nchez-Bl{\'a}zquez} P.,  et~al., 2006, \mn@doi [\mnras]
  {10.1111/j.1365-2966.2006.10699.x}, \href
  {http://adsabs.harvard.edu/abs/2006MNRAS.371..703S} {371, 703}

\bibitem[\protect\citeauthoryear{{Schlegel}, {Finkbeiner}  \&
  {Davis}}{{Schlegel} et~al.}{1998}]{1998ApJ...500..525S}
{Schlegel} D.~J.,  {Finkbeiner} D.~P.,   {Davis} M.,  1998, \mn@doi [\apj]
  {10.1086/305772}, \href {http://adsabs.harvard.edu/abs/1998ApJ...500..525S}
  {500, 525}

\bibitem[\protect\citeauthoryear{{Smee} et~al.,}{{Smee}
  et~al.}{2013}]{2013AJ....146...32S}
{Smee} S.~A.,  et~al., 2013, \mn@doi [\aj] {10.1088/0004-6256/146/2/32}, \href
  {http://adsabs.harvard.edu/abs/2013AJ....146...32S} {146, 32}

\bibitem[\protect\citeauthoryear{{Smith} et~al.,}{{Smith}
  et~al.}{2002}]{2002AJ....123.2121S}
{Smith} J.~A.,  et~al., 2002, \mn@doi [\aj] {10.1086/339311}, \href
  {http://adsabs.harvard.edu/abs/2002AJ....123.2121S} {123, 2121}

\bibitem[\protect\citeauthoryear{{Smith}, {Lucey}, {Price}, {Hudson}  \&
  {Phillipps}}{{Smith} et~al.}{2012}]{2012MNRAS.419.3167S}
{Smith} R.~J.,  {Lucey} J.~R.,  {Price} J.,  {Hudson} M.~J.,   {Phillipps} S.,
  2012, \mn@doi [\mnras] {10.1111/j.1365-2966.2011.19956.x}, \href
  {http://adsabs.harvard.edu/abs/2012MNRAS.419.3167S} {419, 3167}

\bibitem[\protect\citeauthoryear{{Smith} et~al.,}{{Smith}
  et~al.}{2015}]{2015MNRAS.454.2502S}
{Smith} R.,  et~al., 2015, \mn@doi [\mnras] {10.1093/mnras/stv2082}, \href
  {http://adsabs.harvard.edu/abs/2015MNRAS.454.2502S} {454, 2502}

\bibitem[\protect\citeauthoryear{{Stinson}, {Dalcanton}, {Quinn}, {Kaufmann}
  \& {Wadsley}}{{Stinson} et~al.}{2007}]{2007ApJ...667..170S}
{Stinson} G.~S.,  {Dalcanton} J.~J.,  {Quinn} T.,  {Kaufmann} T.,   {Wadsley}
  J.,  2007, \mn@doi [\apj] {10.1086/520504}, \href
  {http://adsabs.harvard.edu/abs/2007ApJ...667..170S} {667, 170}

\bibitem[\protect\citeauthoryear{{Strauss} et~al.,}{{Strauss}
  et~al.}{2002}]{2002AJ....124.1810S}
{Strauss} M.~A.,  et~al., 2002, \mn@doi [\aj] {10.1086/342343}, \href
  {http://adsabs.harvard.edu/abs/2002AJ....124.1810S} {124, 1810}

\bibitem[\protect\citeauthoryear{{Taylor} et~al.,}{{Taylor}
  et~al.}{2015}]{2014arXiv1408.5984T}
{Taylor} E.~N.,  et~al., 2015, \mn@doi [\mnras] {10.1093/mnras/stu1900}, \href
  {http://adsabs.harvard.edu/abs/2015MNRAS.446.2144T} {446, 2144}

\bibitem[\protect\citeauthoryear{{Thomas}, {Brimioulle}, {Bender}, {Hopp},
  {Greggio}, {Maraston}  \& {Saglia}}{{Thomas}
  et~al.}{2006}]{2006A&A...445L..19T}
{Thomas} D.,  {Brimioulle} F.,  {Bender} R.,  {Hopp} U.,  {Greggio} L.,
  {Maraston} C.,   {Saglia} R.~P.,  2006, \mn@doi [\aap]
  {10.1051/0004-6361:200500215}, \href
  {http://adsabs.harvard.edu/abs/2006A%26A...445L..19T} {445, L19}

\bibitem[\protect\citeauthoryear{{Thomas}, {Maraston}, {Schawinski}, {Sarzi}
  \& {Silk}}{{Thomas} et~al.}{2010}]{2010MNRAS.404.1775T}
{Thomas} D.,  {Maraston} C.,  {Schawinski} K.,  {Sarzi} M.,   {Silk} J.,  2010,
  \mn@doi [\mnras] {10.1111/j.1365-2966.2010.16427.x}, \href
  {http://adsabs.harvard.edu/abs/2010MNRAS.404.1775T} {404, 1775}

\bibitem[\protect\citeauthoryear{{Toloba}, {Boselli}, {Cenarro}, {Peletier},
  {Gorgas}, {Gil de Paz}  \& {Mu{\~n}oz-Mateos}}{{Toloba}
  et~al.}{2011}]{2011A&A...526A.114T}
{Toloba} E.,  {Boselli} A.,  {Cenarro} A.~J.,  {Peletier} R.~F.,  {Gorgas} J.,
  {Gil de Paz} A.,   {Mu{\~n}oz-Mateos} J.~C.,  2011, \mn@doi [\aap]
  {10.1051/0004-6361/201015344}, \href
  {http://adsabs.harvard.edu/abs/2011A%26A...526A.114T} {526, A114}

\bibitem[\protect\citeauthoryear{{Toloba} et~al.,}{{Toloba}
  et~al.}{2014}]{2014ApJ...783..120T}
{Toloba} E.,  et~al., 2014, \mn@doi [\apj] {10.1088/0004-637X/783/2/120}, \href
  {http://adsabs.harvard.edu/abs/2014ApJ...783..120T} {783, 120}

\bibitem[\protect\citeauthoryear{{Toloba} et~al.,}{{Toloba}
  et~al.}{2015}]{2015ApJ...799..172T}
{Toloba} E.,  et~al., 2015, \mn@doi [\apj] {10.1088/0004-637X/799/2/172}, \href
  {http://adsabs.harvard.edu/abs/2015ApJ...799..172T} {799, 172}

\bibitem[\protect\citeauthoryear{{Trager}, {Worthey}, {Faber}, {Burstein}  \&
  {Gonz{\'a}lez}}{{Trager} et~al.}{1998}]{1998ApJS..116....1T}
{Trager} S.~C.,  {Worthey} G.,  {Faber} S.~M.,  {Burstein} D.,   {Gonz{\'a}lez}
  J.~J.,  1998, \mn@doi [\apjs] {10.1086/313099}, \href
  {http://adsabs.harvard.edu/abs/1998ApJS..116....1T} {116, 1}

\bibitem[\protect\citeauthoryear{{Trentham} \& {Tully}}{{Trentham} \&
  {Tully}}{2009}]{2009MNRAS.398..722T}
{Trentham} N.,  {Tully} R.~B.,  2009, \mn@doi [\mnras]
  {10.1111/j.1365-2966.2009.15189.x}, \href
  {http://adsabs.harvard.edu/abs/2009MNRAS.398..722T} {398, 722}

\bibitem[\protect\citeauthoryear{{Vazdekis}, {Ricciardelli}, {Cenarro},
  {Rivero-Gonz{\'a}lez}, {D{\'{\i}}az-Garc{\'{\i}}a}  \&
  {Falc{\'o}n-Barroso}}{{Vazdekis} et~al.}{2012}]{2012MNRAS.424..157V}
{Vazdekis} A.,  {Ricciardelli} E.,  {Cenarro} A.~J.,  {Rivero-Gonz{\'a}lez}
  J.~G.,  {D{\'{\i}}az-Garc{\'{\i}}a} L.~A.,   {Falc{\'o}n-Barroso} J.,  2012,
  \mn@doi [\mnras] {10.1111/j.1365-2966.2012.21179.x}, \href
  {http://adsabs.harvard.edu/abs/2012MNRAS.424..157V} {424, 157}

\bibitem[\protect\citeauthoryear{{Wilkinson} et~al.,}{{Wilkinson}
  et~al.}{2015}]{2015MNRAS.449..328W}
{Wilkinson} D.~M.,  et~al., 2015, \mn@doi [\mnras] {10.1093/mnras/stv301},
  \href {http://adsabs.harvard.edu/abs/2015MNRAS.449..328W} {449, 328}

\bibitem[\protect\citeauthoryear{{Wolf} et~al.,}{{Wolf}
  et~al.}{2009}]{2009MNRAS.393.1302W}
{Wolf} C.,  et~al., 2009, \mn@doi [\mnras] {10.1111/j.1365-2966.2008.14204.x},
  \href {http://adsabs.harvard.edu/abs/2009MNRAS.393.1302W} {393, 1302}

\bibitem[\protect\citeauthoryear{{Wolf}, {Martinez}, {Bullock}, {Kaplinghat},
  {Geha}, {Mu{\~n}oz}, {Simon}  \& {Avedo}}{{Wolf}
  et~al.}{2010}]{2010MNRAS.406.1220W}
{Wolf} J.,  {Martinez} G.~D.,  {Bullock} J.~S.,  {Kaplinghat} M.,  {Geha} M.,
  {Mu{\~n}oz} R.~R.,  {Simon} J.~D.,   {Avedo} F.~F.,  2010, \mn@doi [\mnras]
  {10.1111/j.1365-2966.2010.16753.x}, \href
  {http://adsabs.harvard.edu/abs/2010MNRAS.406.1220W} {406, 1220}

\bibitem[\protect\citeauthoryear{{Yan} et~al.,}{{Yan}
  et~al.}{2016}]{2016AJ....151....8Y}
{Yan} R.,  et~al., 2016, \mn@doi [\aj] {10.3847/0004-6256/151/1/8}, \href
  {http://adsabs.harvard.edu/abs/2016AJ....151....8Y} {151, 8}

\bibitem[\protect\citeauthoryear{{York} et~al.,}{{York}
  et~al.}{2000}]{2000AJ....120.1579Y}
{York} D.~G.,  et~al., 2000, \mn@doi [\aj] {10.1086/301513}, \href
  {http://adsabs.harvard.edu/abs/2000AJ....120.1579Y} {120, 1579}

\bibitem[\protect\citeauthoryear{{de Rijcke}, {Michielsen}, {Dejonghe},
  {Zeilinger}  \& {Hau}}{{de Rijcke} et~al.}{2005}]{2005A&A...438..491D}
{de Rijcke} S.,  {Michielsen} D.,  {Dejonghe} H.,  {Zeilinger} W.~W.,   {Hau}
  G.~K.~T.,  2005, \mn@doi [\aap] {10.1051/0004-6361:20042213}, \href
  {http://adsabs.harvard.edu/abs/2005A%26A...438..491D} {438, 491}

\bibitem[\protect\citeauthoryear{{de Zeeuw} et~al.,}{{de Zeeuw}
  et~al.}{2002}]{2002MNRAS.329..513D}
{de Zeeuw} P.~T.,  et~al., 2002, \mn@doi [\mnras]
  {10.1046/j.1365-8711.2002.05059.x}, \href
  {http://adsabs.harvard.edu/abs/2002MNRAS.329..513D} {329, 513}

\bibitem[\protect\citeauthoryear{{van der Marel}, {Alves}, {Hardy}  \&
  {Suntzeff}}{{van der Marel} et~al.}{2002}]{2002AJ....124.2639V}
{van der Marel} R.~P.,  {Alves} D.~R.,  {Hardy} E.,   {Suntzeff} N.~B.,  2002,
  \mn@doi [\aj] {10.1086/343775}, \href
  {http://adsabs.harvard.edu/abs/2002AJ....124.2639V} {124, 2639}

\makeatother
\end{thebibliography}



\appendix
\section{Kinematic maps and profiles}
\label{appen1}

\begin{figure*}
\begin{center}
\includegraphics[width=0.88\textwidth]{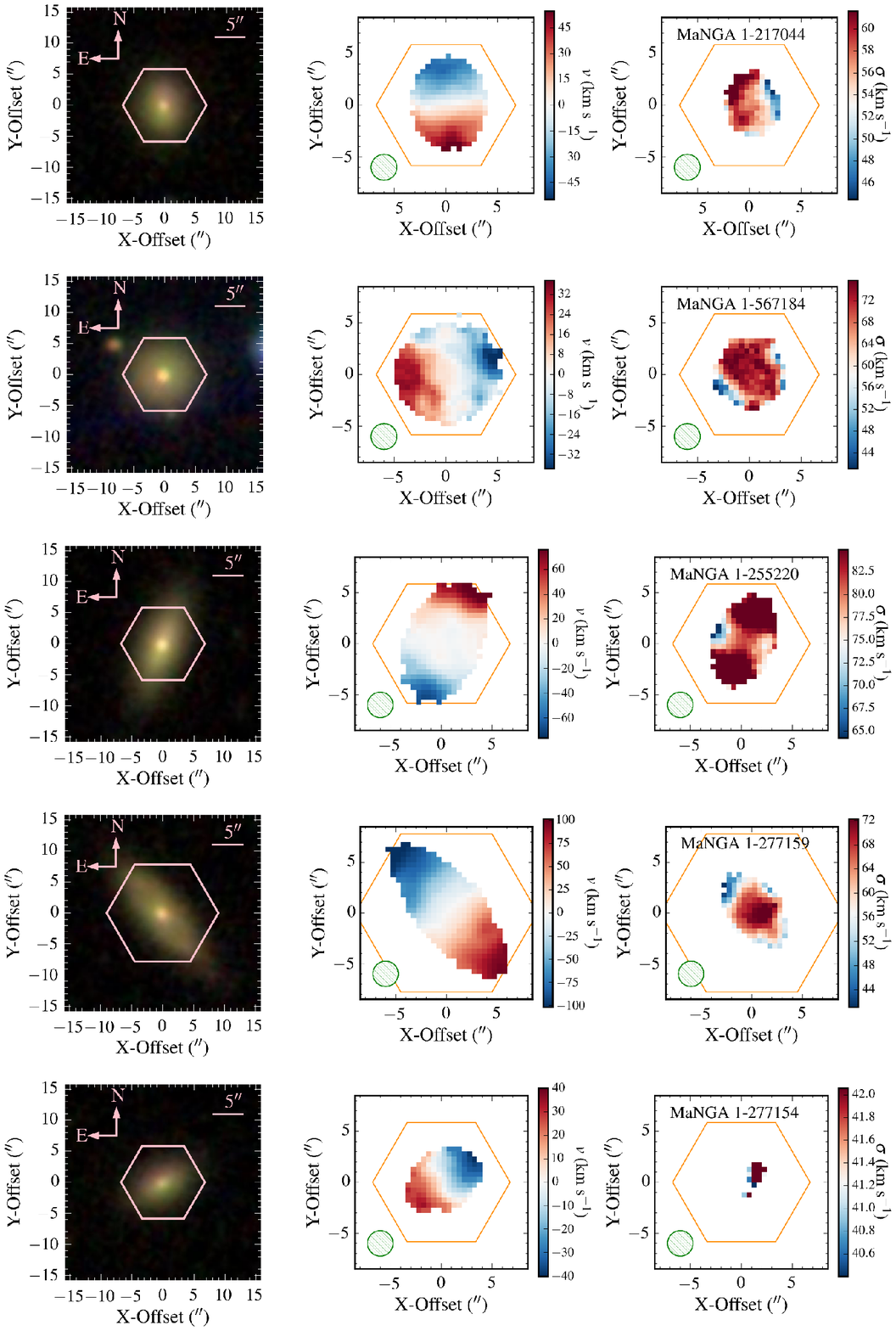}
\caption{SDSS colour images (left panel), radial velocity maps (middle panel), and stellar velocity/$\sigma$ maps for the MaNGA quenched faint galaxies. The stellar velocity maps only include spaxels with $S/N>10$, and the stellar velocity dispersion maps only contain spaxels with $S/N > 20$ and $\sigma_{\star} > 40$~km~s$^{-1}$. The maps are uncorrected for galaxy inclination and beam smearing effects. \label{fig:apmaps}}
\end{center}
\end{figure*}

Kinematic maps are provided in Fig.~\ref{fig:apmaps} for a selection the faint, quenched galaxies examined in this work.  The maps are constructed using data from unbinned datacubes. The stellar velocity maps only include spaxels with $S/N>10$, while the stellar velocity dispersion maps only include spaxels with $S/N > 20$ and $\sigma_{\star} > 40$~km~s$^{-1}$. The maps are uncorrected for galaxy inclination or for the effect of beam smearing. 

\bsp	
\label{lastpage}
\end{document}